\documentclass[twocolumn]{aastex63}

\usepackage{multirow}
\usepackage{natbib}
\usepackage{lineno}

\def\h{\hskip -3.0 mm}
\def\nd{\nodata}

\def\apjl{{ApJL}}
\def\apjs{{ApJS}}
\def\asec{$^{\prime\prime}$}

\def\e#1{$\times$10$^{#1}$}

\def\farcs{\hbox{$.\mkern-4mu^{\prime\prime}$}}

\def\hal{H$\alpha$}
\def\hb{H$\beta$}

\def\hst{{HST}}
\def\swift{{Swift}}

\def\lax{{$\mathrel{\hbox{\rlap{\hbox{\lower4pt\hbox{$\sim$}}}\hbox{$<$}}}$}}
\def\gax{{$\mathrel{\hbox{\rlap{\hbox{\lower4pt\hbox{$\sim$}}}\hbox{$>$}}}$}}
\def\simlt{\lower.5ex\hbox{$\; \buildrel < \over \sim \;$}}
\def\simgt{\lower.5ex\hbox{$\; \buildrel > \over \sim \;$}}
\def\lum{erg s$^{-1}$}
\def\mbh{{$M_{\rm BH}$}}

\def\cm2{cm$^{-2}$}

\def\solmass{$M_\odot$}

\def\lbol{$L_{{\rm bol}}$}
\def\ledd{$L_{{\rm Edd}}$}

\def\mlb{$M_{\rm BH}-L_{\rm{bul}}$}

\def\mm{$M_{\rm BH}-M_{\star}$}

\def\msig{$M_{\rm BH}-\sigma_\star$}

\def\edd{$L_{{\rm bol}}$/{$L_{{\rm Edd}}$}}
\def\lbul{$L_{\rm bul}$}

\def\vel{$\sigma_{\star}$}

\def\nd{\nodata}
\def\h{\hskip -2.0 mm}

\shorttitle{Host Galaxies of Swift-BAT AGN}
\shortauthors{KIM et al.}

\begin{document}

\title{A Hubble Space Telescope Imaging Survey of Low-Redshift Swift-BAT Active Galaxies\footnote{Based on observations made with the NASA/ESA Hubble Space Telescope, obtained at the Space Telescope Science Institute, which is operated by the Association of Universities for Research in Astronomy, Inc., under NASA contract NAS5-26555. These observations are associated with program \#15444.}}

\author{Minjin Kim}

\affiliation{Department of Astronomy and Atmospheric Sciences, 
Kyungpook National University, Daegu 702-701, Korea; mkim.astro@gmail.com}

\author{Aaron J. Barth}

\affiliation{Department of Physics and Astronomy, 4129 Frederick Reines Hall, University of California, Irvine, CA 92697-4575, USA}

\author{Luis C. Ho}

\affiliation{Kavli Institute for Astronomy and Astrophysics, Peking 
University, Beijing 100871, China; lho.pku@gmail.com}

\affiliation{Department of Astronomy, School of Physics, Peking University, Beijing 100871, China}

\author{Suyeon Son}
\affiliation{Department of Astronomy and Atmospheric Sciences,
Kyungpook National University, Daegu 702-701, Korea; mkim.astro@gmail.com}

\begin{abstract}

We present initial results from a Hubble Space Telescope snapshot imaging survey of the host galaxies of \swift-BAT active galactic nuclei (AGN) at $z<0.1$. The hard X-ray selection makes this sample sample relatively unbiased in terms 
of obscuration compared to optical AGN selection methods. 
The high-resolution images of 154 target AGN enable
us to investigate the detailed photometric structure of the host galaxies, 
such as the Hubble type and merging features. We find that 48\% and 44\% of 
the sample is hosted by early-type and late-type galaxies, respectively. 
The host galaxies of the remaining 8\% of the sample are classified as
peculiar galaxies because they are heavily disturbed.
Only a minor fraction of host galaxies (18\%$-$25\%) exhibit merging features
(e.g., tidal tails, shells, or major disturbance). 
The merging fraction increases strongly as a function of bolometric AGN luminosity, revealing that merging plays an important 
role in triggering luminous AGN in this sample. However, the merging fraction is weakly 
correlated with the Eddington ratio, suggesting that merging does not
necessarily lead to an enhanced Eddington ratio. Type~1 and type~2 AGN are 
almost indistinguishable in terms of their Hubble type distribution and merging fraction. 
However, the merging fraction of type~2 AGN peaks at a lower bolometric luminosity 
compared with those of type~1 AGN. This result may imply that the triggering mechanism 
and evolutionary stages of type~1 and type~2 AGN are not identical. 
\end{abstract}


\section{Introduction} 

In massive galaxies that exhibit bulges, supermassive black holes (SMBHs) are 
ubiquitous. The strong correlations between SMBH mass and the physical properties 
(e.g., luminosity, velocity dispersion, and stellar mass) of the spheroidal 
component of the host galaxy imply a causal connection between 
a SMBH and its host galaxy in terms of their formation and evolution
(\citealt{magorrian_1998, gebhardt_2000, ferrarese_2000, kormendy_2013}). 
According to theoretical studies, active galactic nuclei (AGN) that originate
from accreting SMBH play an important role in regulating BH growth and star
formation, which, in turn, can produce the strong correlations between SMBHs and 
the properties of their host galaxies (\citealt{fabian_1999, dimatteo_2005, hopkins_2005}). 
In this model, SMBH growth and star formation occur primarily 
during the AGN phase, which can be triggered by gas-rich mergers or interactions with 
companion galaxies (\citealt{sanders_1988, hopkins_2008}). Therefore, identifying the 
merging features of AGN hosts is vital to understand the galaxy$-$host connection.

According to the AGN unification model, the accretion disk
and broad-line region are directly observed in type~1 AGN, while those 
central structures are visually obscured by the sub-parsec dusty torus in type~2 AGN
(\citealt{antonucci_1993,urry_1995}). On the contrary, if an AGN is mainly 
triggered by gas-rich merging, then the gas and dust content in the central regions of the host can be enhanced during the early stages of the merger 
(\citealt{blecha_2018}). A late stage then occurs,
in which the AGN-driven outflow can expel the surrounding gas and dust so that
the AGN is eventually observed to be type~1 (\citealt{sanders_1988, 
ishibashi_2016}). Therefore, the nucleus can be naturally obscured due to the 
dense material in the host galaxy on spatial scales larger than the dusty torus. According to this evolutionary sequence, the host galaxies of type~2 AGN are 
more likely to be disturbed than those of type~1 AGN, or might show an 
association with earlier merger stages. 
On the other hand, \citet{elitzur_2014} argued that type~1 AGN can 
evolve to intermediate-type (1.8-1.9), and finally to type~2 AGN as luminosity decreases.
In this light, investigating the 
physical properties of host galaxies is crucial not only to determine the AGN 
triggering mechanism but also to test AGN unification models. 
Host galaxy properties have been used to test unified models for many 
AGN samples 
(e.g., \citealt{simcoe_1997}). Several studies have found that the merging 
fraction for obscured AGN is higher than that for unobscured AGN 
(e.g., \citealt{urrutia_2008, glikman_2015, donley_2018}). Additionally,
Compton-thick AGN appear to be preferentially hosted by merging galaxies
(\citealt{kocevski_2015, ricci_2017a, koss_2018}). \citet{koss_2011},
however, argued that the properties of the host galaxies of type~1 
and type~2 AGN are indistinguishable. 

Numerous studies have investigated the properties of host galaxies 
of nearby and distant AGN. One of the primary goals of those studies is to 
identify the merging features in the AGN hosts as a possible origin of AGN 
triggering. Several studies have claimed that nearby AGN are preferentially 
hosted by galaxies 
that exhibit merging features compared with normal galaxies 
(\citealt{koss_2011, hong_2015, goulding_2018, ellison_2019, marian_2020}). 
In addition, luminous AGN are more likely to be associated with merging or 
interaction than less luminous AGN (e.g., \citealt{treister_2012,kim_2017}). 
However, some observational studies have found that there is no evidence of an
excess of merging fraction in various types of luminous AGN 
(e.g., \citealt{dunlop_2003,cisternas_2011, bohm_2013, villforth_2014, 
villforth_2017}). 
For example, \citet{zhao_2019} argued that only a minor fraction
Type~2 quasar hosts (34\%) are mergers, suggesting that major 
merging plays only a limited role in triggering AGN. For narrow-line Seyfert 1
(NLS1) galaxies, the fraction of barred hosts is significantly higher than the 
bar fraction of ordinary Type~1 AGN, implying that this subclass is 
predominantly triggered by internal secular evolution rather than by merging 
(\citealt{crenshaw_2003, deo_2006, sani_2010, orban_2011, kim_2017}).    

\begin{figure}[h]
\centering
\includegraphics[width=0.45\textwidth]{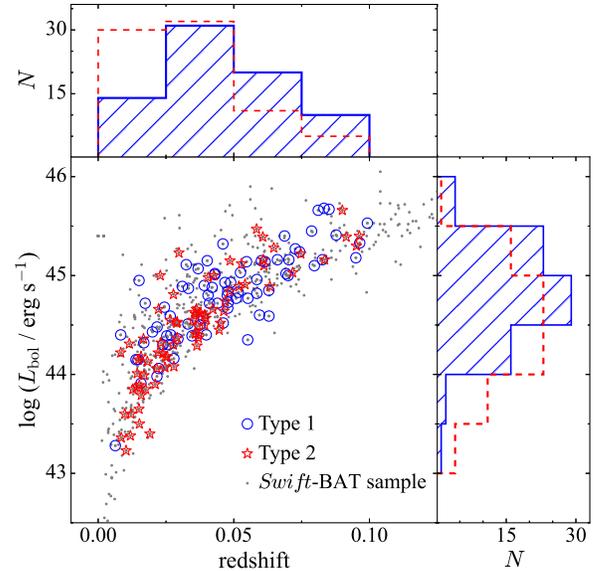}
\caption{
Redshift and bolometric luminosity distributions of the parent sample of
\swift-BAT AGN (small gray dots). The Type~1 AGN observed for this HST program are denoted by blue stars and blue histograms, while Type~2 objects are given by red circles and red histograms.  Bolometric luminosity is estimated from the intrinsic X-ray luminosity (Ricci et al. 2017) assuming a bolometric correction of 8. 
}
\end{figure}

Finally, the photometric properties of host galaxies have been widely used to 
explore the coevolution between SMBHs and host galaxies. The stellar 
populations of AGN host galaxies can provide a useful constraint on the 
SMBH-galaxy connection (e.g., 
\citealt{sanchez_2004, canalizo_2013, matsuoka_2014, kim_2019,zhao_2019}). 
Some studies have shown that host galaxies of type~1 AGN tend to be bluer or 
overluminous compared with inactive galaxies due to the former's young stellar 
population (e.g., \citealt{sanchez_2004, kim_2019}).
However, other investigations have found no sign of young stellar populations 
in AGN hosts (e.g., \citealt{nolan_2001, bettoni_2015}). Additionally, 
the spheroid luminosity of type~1 AGN hosts has also been used to investigate 
the relation between BH mass (\mbh) and bulge luminosity (\lbul) 
(e.g., \citealt{kim_2008b, greene_2010, park_2015, kim_2019, li_2021}).

The aforementioned studies have employed widely different sample selection 
methods and datasets with diverse properties in terms of wavelength or filter 
bands, spatial resolution, and depth, and these differences may be largely 
responsible for the sometimes divergent conclusions reached in different 
investigations. Thus, despite the large observational effort invested into AGN 
host galaxy studies over many years, further work based on large, uniformly 
selected samples with high-quality imaging data can still provide new insights.
Indeed, there have been several attempts to conduct imaging 
surveys of nearby Seyferts using HST (e.g., \citealt{nelson_1996, malkan_1998, 
schade_2000}). Those surveys used early HST images with a relatively 
narrow field of view that did not always cover the full host galaxy 
(the Wide Field Planetary Camera 1 and 2), and early snapshot programs with 
these cameras obtained relatively shallow depth. Images of Type 1 AGN are 
often saturated due to the bright nuclei, which makes it difficult to obtain a 
robust decomposition to determine bulge properties accurately. Furthermore, 
some of these studies \citep[e.g.,][]{malkan_1998} observed samples selected 
from highly heterogeneous criteria based on AGN catalogs available at the time.

To overcome these limitations, we have used HST to conduct an imaging survey 
of nearby AGN at $z<0.1$ selected from the \swift-BAT AGN catalog. \swift-BAT 
AGN are identified from hard X-ray data 
that, compared with optical and infrared data, is less affected by extinction 
and contamination from star formation. Therefore, hard X-ray data provides a
relatively uniform and homogeneous sample in terms of obscuration, with the 
exception of very highly obscured (Compton-thick) AGN. 
The deep and coherent high-resolution imaging dataset from \hst\ allows 
us to robustly investigate the structural properties of the host galaxies for a large and uniformly selected sample. 
Moreover, the Swift-BAT AGN sample has the benefit of extensive multiwavelength measurements 
(e.g., BH mass, bolometric luminosity, Eddington ratio, hydrogen column 
density, and optical classification) that have been carried out in numerous 
studies (\citealt{koss_2017,oh_2017,ricci_2017}). Hence, our imaging survey is 
an excellent complement to the existing body of data for this benchmark AGN 
sample, providing an excellent dataset to study the detailed properties of 
nearby AGN hosts and to investigate the relationships between AGN and host 
properties while minimizing sample selection biases.

\begin{figure}[t!]
\centering
\includegraphics[width=0.45\textwidth]{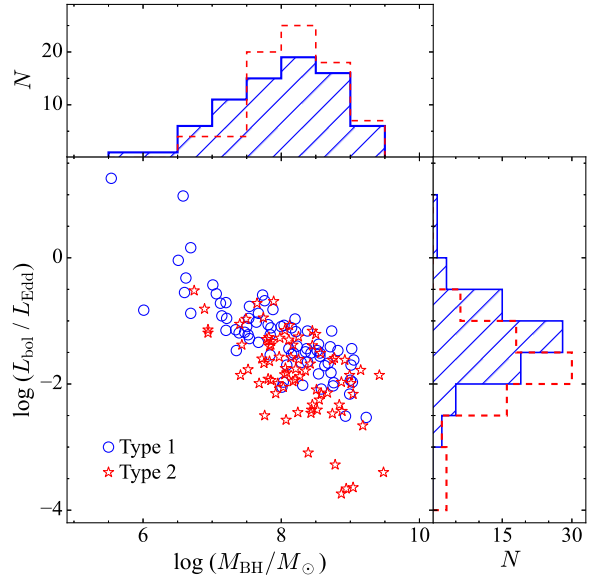}
\caption{
The distributions of BH mass and Eddington ratio of the sample. \mbh\ is measured using the virial method, the \msig\ relation, or the \mm\ relation. 
}
\end{figure}

This paper presents a description of the morphological properties of the host 
galaxies of these low-$z$ \swift-BAT AGN using newly acquired \hst\ 
$I$-band images for 154 objects. Our primary goal is to explore the connection 
between the physical properties of AGN and those of their hosts.  Section~{2} 
describes the sample selection, our imaging data, and the physical properties 
of the sample.  Section~{3} presents visual morphological 
classifications.  We discuss the correlation between the morphological 
structures of the hosts and the properties of the AGN in Section~{4}.  
A summary is given in Section~{5}.  This work adopts the following 
cosmological parameters: $H_0 = 100 h = 67.8$ km s$^{-1}$ Mpc$^{-1}$, $\Omega_m = 0.308$, and $\Omega_{\Lambda} = 0.692$ (\citealt{planck_2016}).

\section{Sample and Data}
\subsection{Sample Selection}

In 2017, the Space Telescope Science Institute announced a special mid-cycle 
call proposals for ``gap-filler'' snapshot surveys to be conducted using the 
Advanced Camera for Surveys (ACS) Wide-Field Camera (WFC). The programmatic 
goal for this call was to provide large, full-sky catalogs of snapshot targets 
that would be used to fill HST scheduling gaps, at lower priority than 
standard general observer and snapshot programs. This survey 
(HST program 15444) was selected as one of three gap-filler programs in the 
2017 call. The proposal call dictated that the total duration of snapshot 
visits should be $\lesssim26$ minutes, which is sufficient to allow for two 
full-frame ACS/WFC exposures with a total on-source exposure time of 
$\sim700$ s after acquisition and other overheads.

Our sample is drawn from the 70-month Swift-BAT X-ray source catalog 
(\citealt{Baumgartner_2013}). Since our primary science goals involve 
morphological studies of AGN host galaxies, we focus on low-redshift objects, 
and we selected Swift-BAT AGN at $z<0.1$. After cross-matching this catalog 
with the HST archive, we excluded objects already having images in an $I$-band 
equivalent filter (in most cases F814W) with the WF/PC2, ACS, or WFC3 cameras, 
to avoid duplications. The remaining set of 543 objects provides an ideal 
gap-filler sample with a large number of targets widely distributed over the 
full sky. No cuts to the sample were made based on Galactic latitude or 
foreground extinction, because HST $I$-band imaging could in principle provide 
new morphological classifications and host galaxy structural information even 
for targets in fields with moderate to high extinction that otherwise might 
not be observed in other surveys.  Figure 1 shows the distribution of 
redshifts and bolometric luminosities for the parent Swift-BAT sample and for 
the set of 543 targets making up the sample for program 15444.

\begin{figure*}[t!]
\centering
\includegraphics[width=1.0\textwidth]{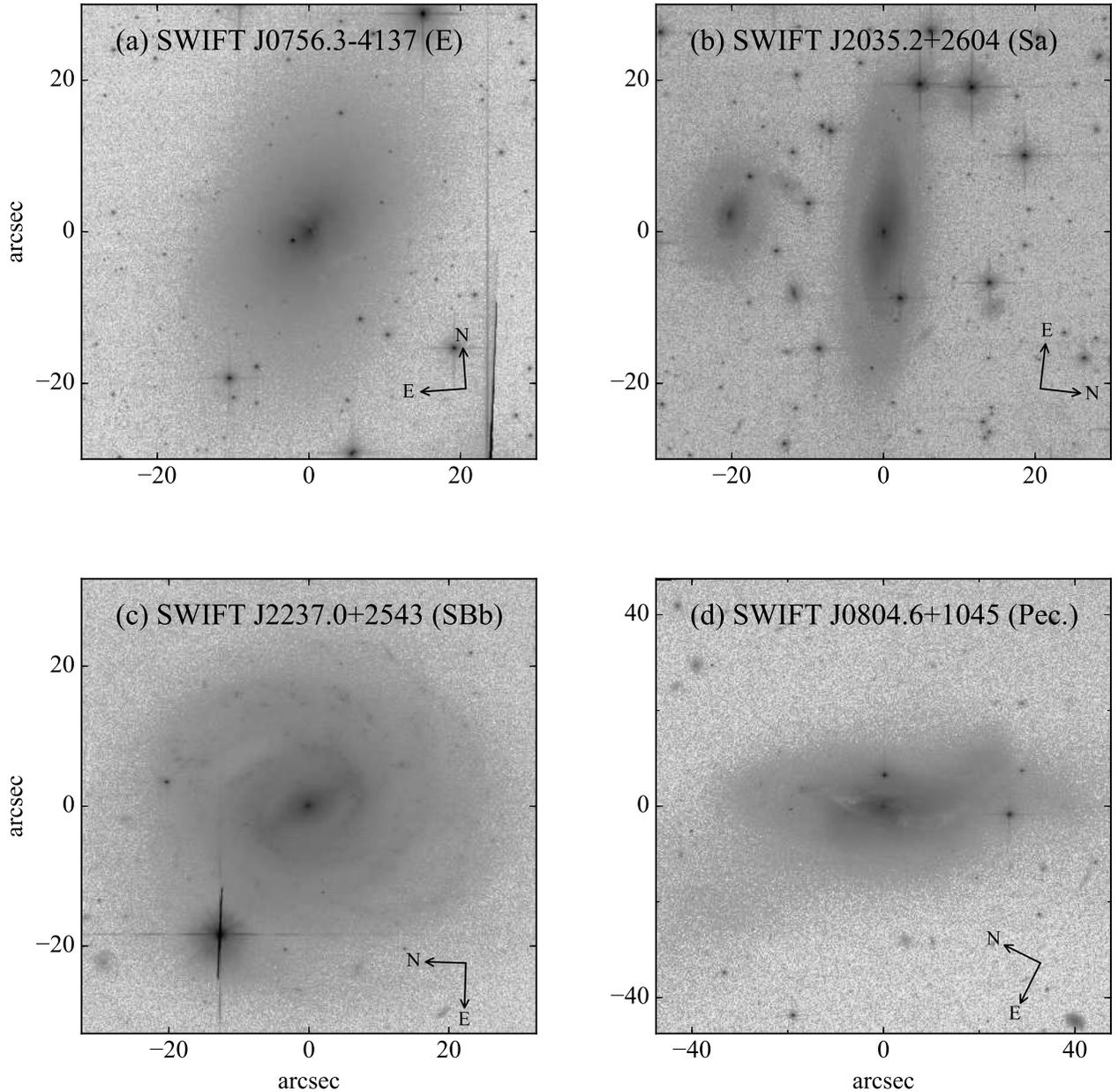}
\caption{
Representative examples of various morphologiccal types: (a) elliptical, (b) Sa spiral, (c) barred SBb spiral, and (d) peculiar.  All images are displayed on an asinh stretch.
}
\end{figure*}

\subsection{Observations}
Observations for Program 15444 began in 2018 March and are ongoing 
(at a low rate) during the first half of 2021. This paper includes data taken 
through 2021 June. We used ACS/WFC with the F814W filter, which is 
approximately equivalent to the $I$ band, to maximize the brightness contrast 
between the host galaxy and the active nucleus. Suffering less dust 
attenuation than filters at shorter wavelengths, F814W also allows us to 
better probe the photometric substructures of the hosts.  We obtained two 
dithered exposures per target with integration times of 337 s, the minimum 
time required to avoid buffer dump overheads. The AGN was centered on the WFC1 
detector. For Type 1 (unobscured) AGN, a nuclear point source is generally 
expected to be present at optical wavelengths \citep{nelson_1996}. 
Accordingly, we took an additional 5 s exposure with the WFC1-B512 subarray to 
obtain an unsaturated image of the nuclear point source and its immediate 
surroundings for objects listed as Type 1 AGN by \citet{Baumgartner_2013} or 
\citet{koss_2017}.  Of the 157 AGN targeted for observations, useful data were 
obtained for 154 sources (Table~1), while three observations were lost due to 
guide star acquisition failures. For other targets, delayed guide star 
acquisition led to slightly shorter exposure times than the planned 
$2\times337$ s (SWIFT~J0533.9$-$1318, SWIFT~J0747.5+6057, and 
SWIFT~J0753.1+4559).  The set of 154 successfully observed targets includes 
75 Type 1 and 79 Type 2 AGN.

\begin{figure*}[t!]
\centering
\includegraphics[width=1.0\textwidth]{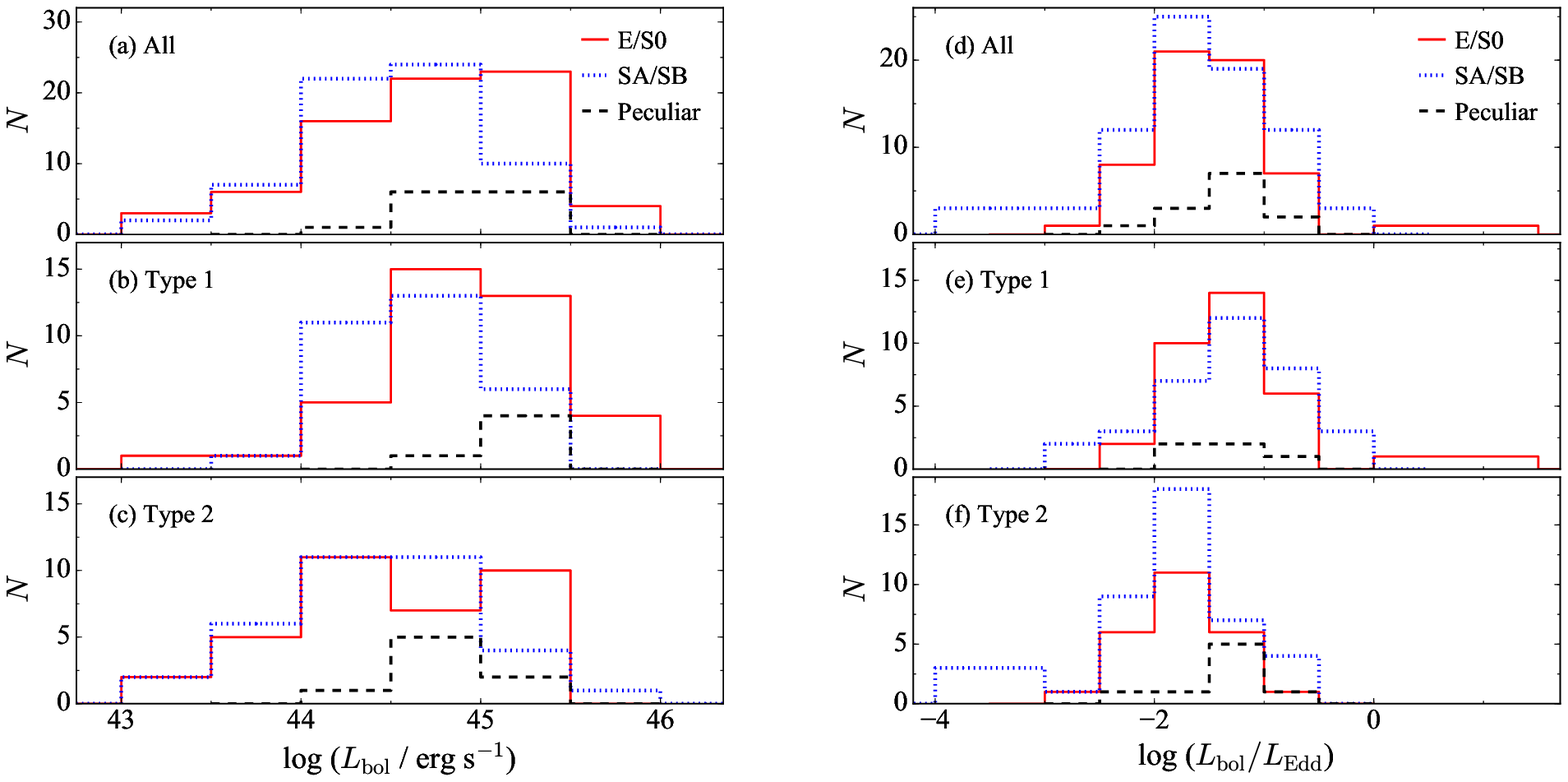}
\caption{
The distributions of bolometric luminosity (\lbol) and Eddington ratio 
(\lbol/\ledd) for different morphological classifications. The red, blue, and 
black histograms denote early-type galaxies (E/S0), late-type galaxies with 
spiral arms (SA/SB), and peculiar galaxies, respectively. 
}
\end{figure*}

\subsection{Data Reduction}
For the long exposures, we adope the standard reduction provided by the 
pipeline, including bias and dark subtraction, flat-field correction, and 
correction for charge-transfer efficiency. 
The cosmic ray rejection was performed when combining the two dithered images 
is imperfect. Therefore, we additionally applied the {\tt LACosmic} algorithm 
(\citealt{vandokkum_2001}) to the individual images to improve the cosmic ray 
rejection.  We used the PyRAF task {\scshape TweakReg} to determine a robust 
shift between the dithered images.  The final combined images, generated with 
{\scshape AstroDrizzle}, have a pixel scale of $\sim$0\farcs05 and a 
field-of-view of $\sim 202$\asec$\times202$\asec.  We measured the variance of 
the sky values after masking all objects to determine the depth of the images, 
which on average we estimate to be $\sim23.6$ mag~arcsec$^{-1}$ ($1\,\sigma$). 

The short-exposure sub-array images for the type~1 objects exhibit horizontal 
bands introduced by bias striping noise (\citealt{grogin_2010, grogin_2011}),
which were removed separately with the {\scshape acs\_destripe\_plus task}.  
The saturated pixels in the combined long-exposure image were replaced with 
the same pixels in the short, unsaturated image. 
The short and long exposures 
were aligned using the position of the nucleus. The central position of the 
nucleus for the long exposures was determined after masking the saturated 
pixels.

\subsection{Physical Properties of AGN}

To investigate the connection between AGN and their host galaxies, we 
assembled the optical spectral properties of the sample from the literature to 
estimate physical properties such as BH mass, accretion rate, and bolometric 
luminosity. For type~1 AGN, we use the virial method 
($M_{\rm BH} \propto Rv^2/G$) to estimate BH mass by combining the size 
($R_{\rm BLR}$) and velocity dispersion ($v$) of the broad-line region (BLR).  
The BLR size can be estimated through reverberation mapping experiments 
(\citealt{blandford_1982}) and from single-epoch optical spectroscopy using 
the empirical relation between $R_{\rm BLR}$ and AGN luminosity 
(\citealt{kaspi_2000,bentz_2013}).  The intrinsic scatter of the BLR 
size-luminosity relation (\citealt{du_2018, alvarez_2020}) 
introduces significant systematic uncertainty 
into the BH masses derived from single-epoch spectra. 
For single-epoch mass estimates, the assumption of a single virial scaling factor contributes additional error to the mass estimates, since the true virial factor for a given AGN will depend on the geometry and kinematics of the BLR. 
While the mean virial factor has been determined by assuming that AGN follow the 
same \msig\ relation as the inactive galaxies, its calibration appears to be sensitive 
morphologies of galaxies and types of AGNs \citep[e.g.,][]{ho_2014}. 
Here, we 
recalculate the BH masses rather than using the values from \citet{koss_2017}.
We adopt the virial 
mass estimator based on \hb\ and 5100 \AA\ AGN continuuum luminosity as 
calibrated by \citet{ho_2015} for all bulge types, using spectral measurements 
from \citet{koss_2017} of 56 type~1 AGN.  We use the scaling relations of 
\citet{greene_2005} to convert \hal\ and \hb\ luminosities to 5100 \AA\ AGN 
luminosity, and we scale broad \hal\ line widths and broad \hb\ line widths.

\begin{figure*}[t!]
\centering
\includegraphics[width=1.0\textwidth]{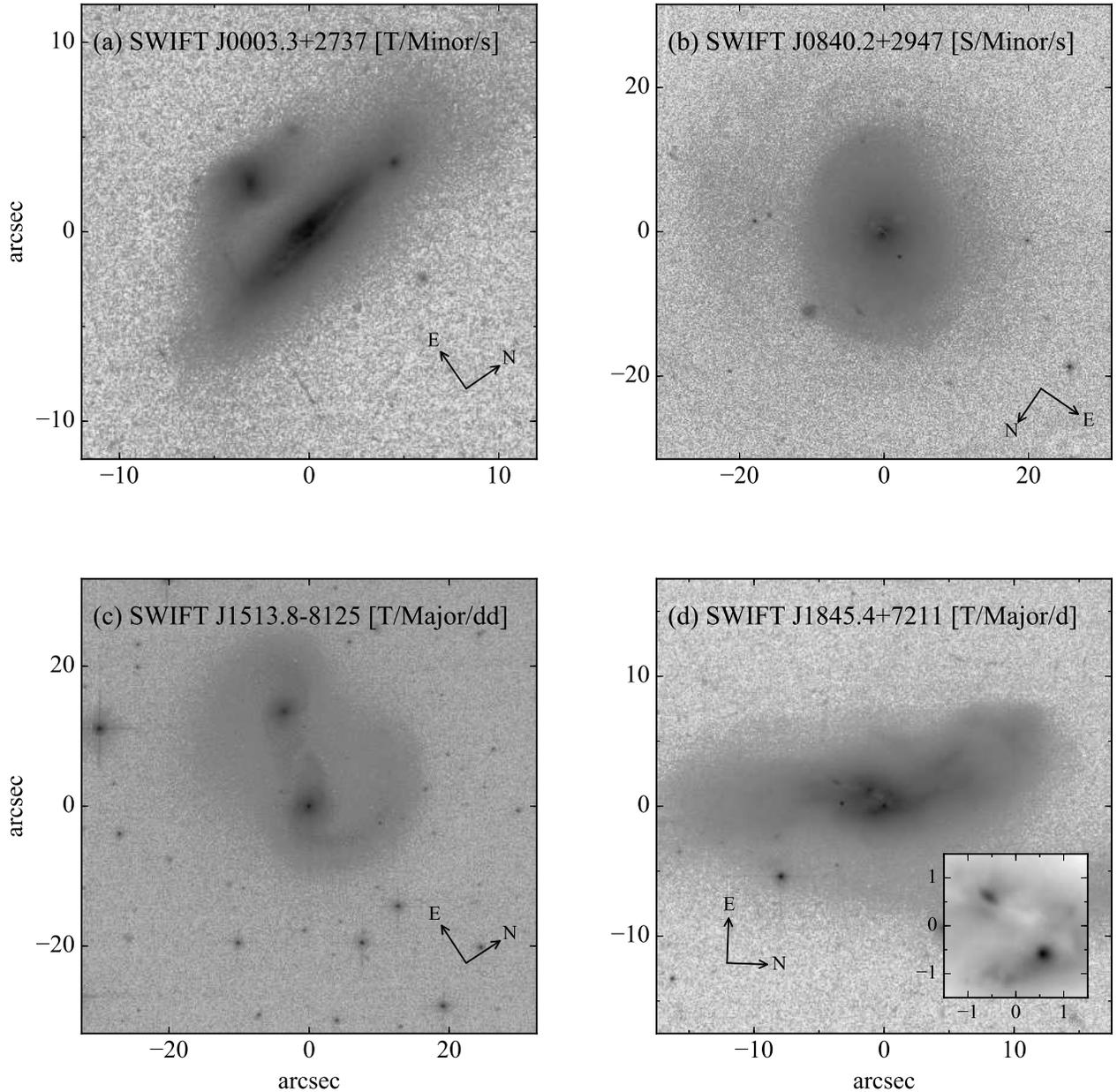}
\caption{
Example images of subgroups based on the various merging properties: (a) minor merger with tidal tail and single nucleus; (b) minor merger with shell and single nucleus; (c) major merger with tidal tail and double nuclei in separate hosts; and (d) major merger with tidal tail and double nuclei in a single host.  In panel (d), a close-up of the central region is shown in the inset.
}
\end{figure*}

\begin{figure*}[t!]
\centering
\includegraphics[width=1.0\textwidth]{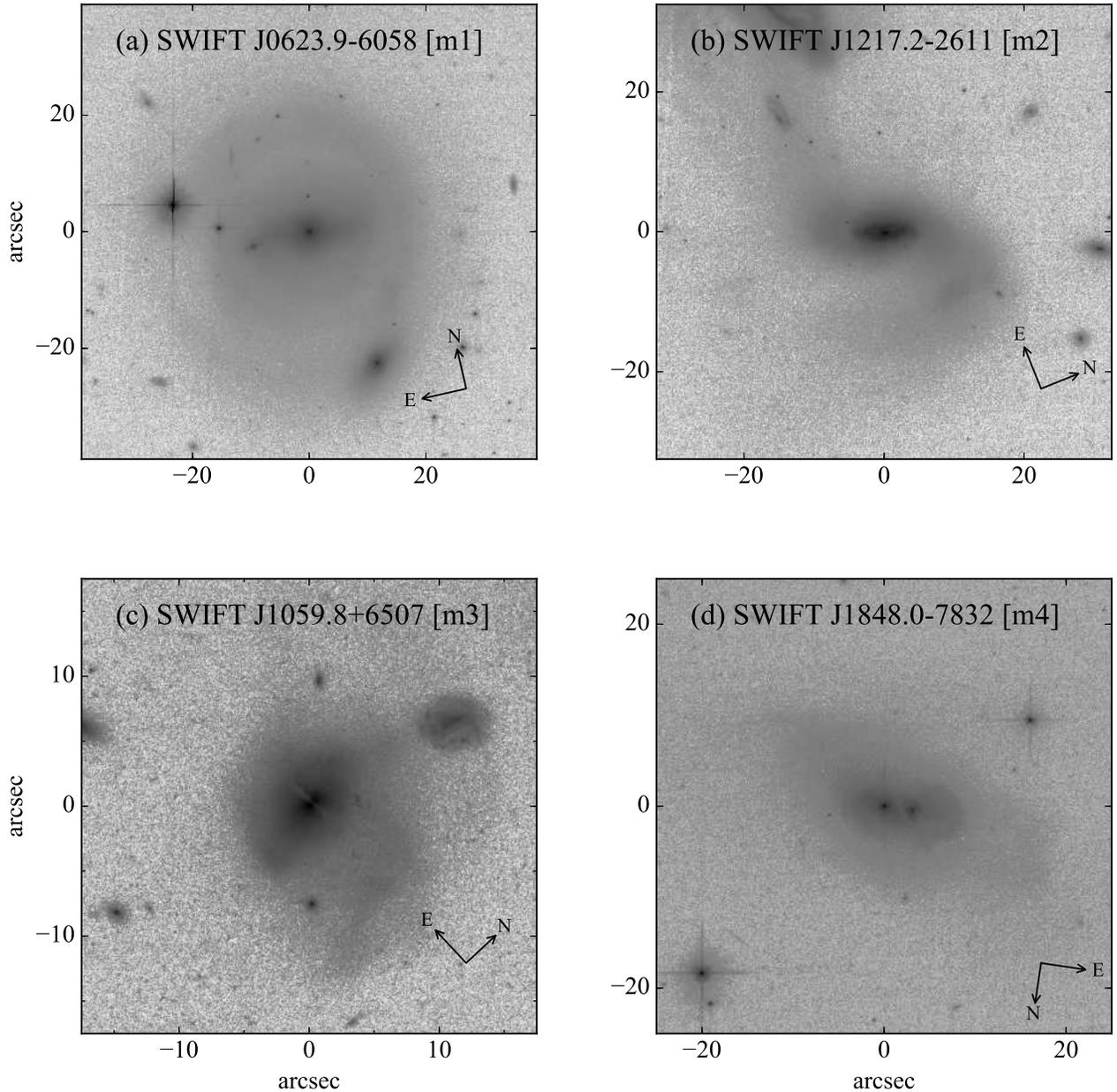}
\caption{
Example images of different subgroups based on their merging stages: (a) accompanying a galaxy with no (or weak) sign of interaction (m1); (b) merging galaxy without the major disturbance of the host (m2); (c) major disturbance but the companion galaxy is not yet merged (m3); (d) merging host with two galaxies sharing a common envelope (m4).  }
\end{figure*}

For the Type~1 AGN lacking useful measurements of broad emission lines but 
that do have available stellar velocity dispersion (\vel) measurements, and 
for Type 2 objects having \vel\ available, we compute \mbh\ using the \msig\ 
relation from \citet{kormendy_2013}\footnote{We use the \msig\ relation for 
classical bulges and elliptical galaxies: 
$\log (M_{\rm BH}/M_\odot)= 8.49 + 4.38 \log (\sigma_*/200\,{\rm km~s^{-1}})$.}.

For the remaining 63 objects without broad-line measurements or information on 
\vel, we estimate \mbh\ from the total stellar mass of the host galaxy 
($M_\star$). Although the $M_{\rm BH}-M_\star$ relation is not as tight as the 
\msig\ relation, it can still yield approximate BH masses.  The \mm\ relation 
is sensitive to the host galaxy morphological type. For the sake of simplicity,
we adopt the \mm\ relation for all galaxy types, which has an uncertainty of 
$\sim0.8$~dex (\citealt{greene_2020})\footnote{We use  the \mm\ relation for 
all galaxy types: $\log (M_{\rm BH}/M_\odot) = 7.43 + 1.61 \log 
({M_\star}/{3\times10^{10}M_\odot})$.}.  We obtain the integrated F814W-band 
flux of the host galaxy through elliptical isophotal fitting of the observed 
images using the {\tt IRAF} task {\tt ellipse}, after masking blended nearby 
galaxies and foreground stars using {\tt SExtractor} (\citealt{bertin_1996}). 
For type~1 AGN, the nucleus is subtracted prior to isophotal fitting using the 
point-spread function generated from isolated stars in the science images 
(S. Son et al., in preparation). The $I$-band flux is corrected for Galactic 
extinction (\citealt{schlafly_2011}), and a $k$-correction is applied 
according to the method described by  \citet{chilingarian_2010} using host 
galaxy $V-I_{\rm C}$ colors assigned based on the morphological classification 
(Section~4.1; \citealt{Fukugita_1995}). Along with the mass-to-light ratio 
inferred from the $V-I_{\rm C}$ color (\citealt{into_2013}), the integrated 
$I$-band luminosity then gives $M_\star$.

Note that assigning a fixed color based on galaxy morphology likely 
overestimates the true stellar mass if AGN host galaxies preferentially 
contain a younger stellar population (\citealt{sanchez_2004,zhao_2019}), 
especially in systems with higher Eddington ratio (Kim \& Ho 2019).  However, 
apart from later-type spirals with pseudo bulges (\citealt{zhao_2021}), the 
effect of young stars may not be substantial in the $I$ band.  Pseudo bulges 
contribute $\lesssim 20\%$ to the total light in disk galaxies 
(e.g., \citealt{gao_2019}).  This, in conjunction with the fact that our sample 
has relatively low Eddington ratios (median 
$\langle \log L_{\rm bol}$/$L_{\rm Edd}\rangle =-1.5$), leads us to suspect 
that the systematic uncertainty due to mass-to-light ratio should be small 
compared with the intrinsic scatter in the \mm\ relation. 
 
We use the intrinsic hard X-ray ($14 - 195$~keV) luminosity derived from 
spectral fitting of the \swift-BAT data by \citet{ricci_2017} to estimate the 
bolometric luminosity using a conversion factor of 8 
(\citealt{koss_2017,ricci_2017})\footnote{X-ray measurements are unavailable 
for SWIFT~J1737.7$-$5956A.}. The AGN bolomteric luminosities range 
from $L_{\rm bol} = 10^{43.2} - 10^{45.7}\, {\rm erg~s^{-1}}$, and, for 
Eddington luminosity 
$L_{\rm Edd} = 1.26\times10^{38} (M_{\rm BH}/M_{\odot})\,{\rm  erg~s^{-1}}$, 
the Eddington ratios span  \edd\ $= 0.00018-18.2$.  Finally, we take the
column density of neutral hydrogen ($N_{\rm H}$) deduced from spectral 
analysis of the X-ray data, as assembled by \citet{ricci_2017}.  A lower limit 
of $N_{\rm H} \approx 10^{20}$ \cm2\ yields negligible obscuration along the 
line of sight.
The distributions of BH mass and Eddington ratio are displayed in Figure~2.
The comparison reveals that the BH masses of type~2 AGN are systematically 
higher than those of type~1 by $\sim0.16$ dex. However, given the 
uncertainties of BH mass measurements
($0.3-0.8$ dex) and the different methods applied for the two populations, 
it is unclear if this represents a genuine difference between the type 1 
and type 2 subsamples.

Finally, we also estimate the axis ratio ($b/a$, where $b$ is the semi-minor 
axis and $a$ is the semi-major axis) of the host galaxy using the results from 
the elliptical isophotal analysis.

\section{Analysis and Results}

\subsection{Morphological Classification}
All the targets are classified into approximate morphological types, based on 
visual inspection independently conducted by three of the authors (MK, AB, LH). 
We assign the morphologies into four basic Hubble types: elliptical (E), 
lenticular (S0), spiral (S), and peculiar. The disk galaxies are additionally 
scrutinized for the possible presence of a bar (SA/SB) and the relative 
prominence of their bulge and disk (Sa: bulge-dominated; Sb: intermediate 
bulge; Sc: disk-dominated). Where disagreement arises among the three 
classifiers, they reinspect the images until they reach convergence.  The 
classifications of the sample are summarized in Table~2. The sample is roughly 
evenly divided between early-type hosts (74 or 48\% E/S0) and late-type hosts 
(67 or 44\% spirals), with peculiar morphologies accounting for the minority 
(13 or 8\%).  Representative examples of each Hubble type are given in Figure~3.
 
Figure~4 shows the distributions of Hubble type as a function of bolometric 
luminosity and Eddington ratio. The ellipticals and peculiar galaxies appear 
to be found preferentially in more luminous AGN. However, this trend is less 
significant in Eddington ratio. We also find that there is no substantial 
difference in the Hubble type distribution of type~1 versus type~2 AGN. 

\begin{figure*}[t!]
\centering
\includegraphics[width=1.0\textwidth]{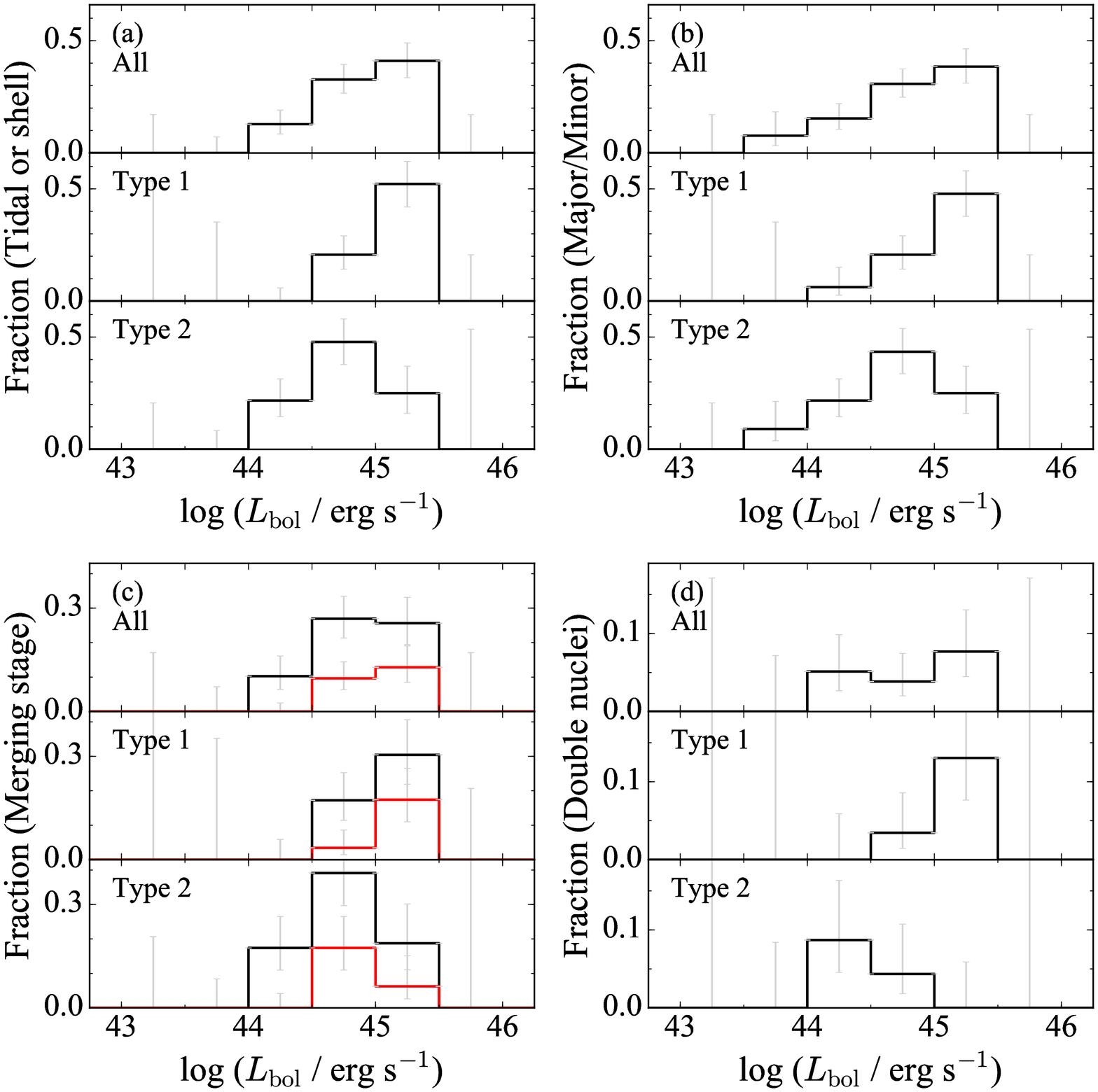}
\caption{
Dependence of the merging fraction on bolometric luminosity, for hosts with (a) a tidal tail or shell; (b) a major or minor merger; (c) tidal features that belong to merging stages ``m2'', ``m3'', or ``m4'' (black histograms) and those in the latest merging stages (``m4''; red histogram); (d) fraction of hosts having double nuclei (``d'' or ``dd'').  Each panel shows separate distributions for all AGN, type~1 AGN, and type~2 AGN.
}
\end{figure*}

\begin{figure*}[t!]
\centering
\includegraphics[width=1.0\textwidth]{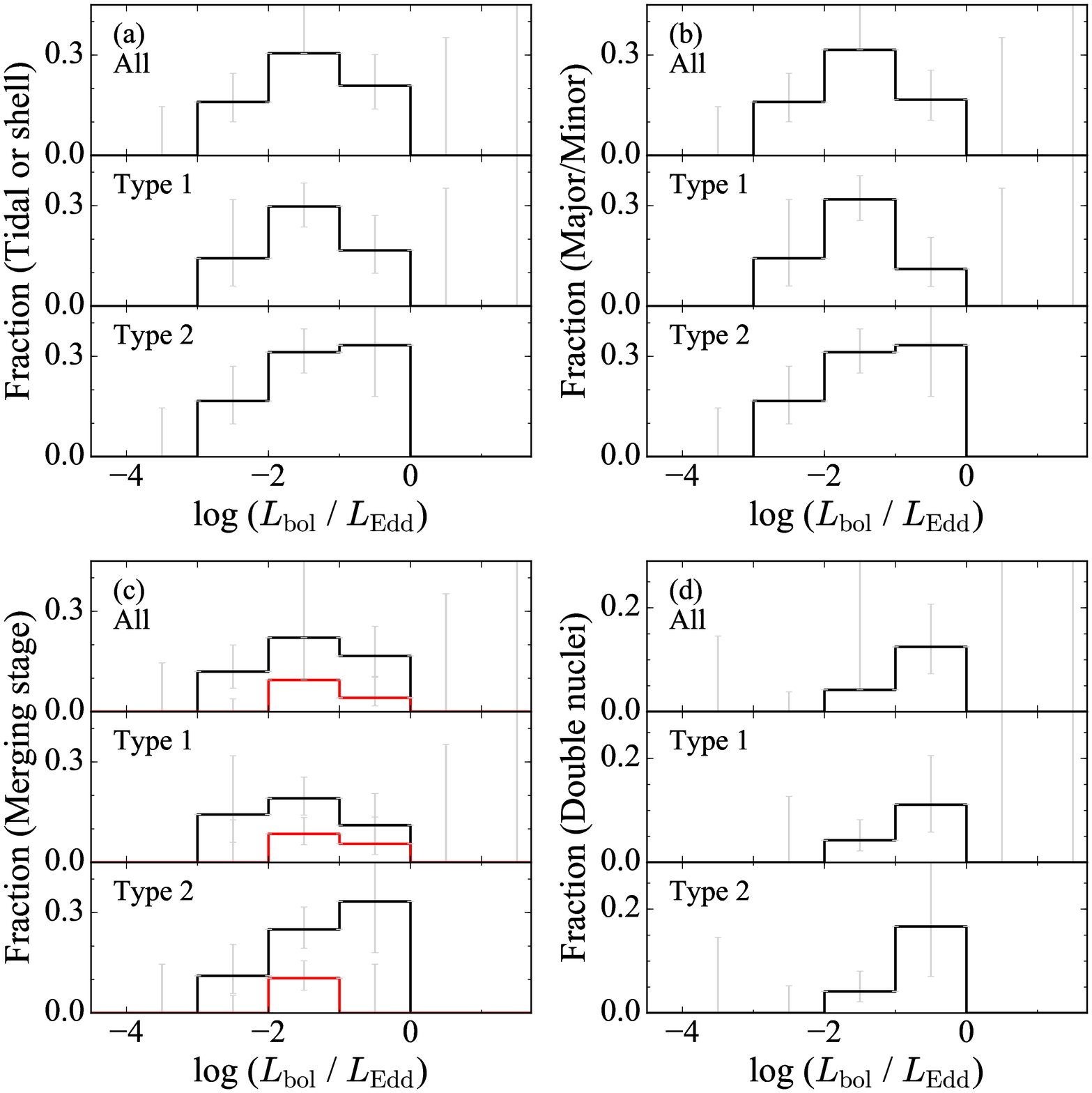}
\caption{
Dependence of the merging fraction on Eddington ratio, for hosts with (a) a tidal tail or shell; (b) a major or minor merger; (c) tidal features that belong to merging stages ``m2'', ``m3'', or ``m4'' (black histograms) and those in the latest merging stages (``m4''; red histogram); (d) fraction of hosts having double nuclei (``d'' or ``dd'').  Each panel shows separate distributions for all AGN, type~1 AGN, and type~2 AGN.
}
\end{figure*}

\subsection{Signatures of Dynamical Perturbations}

In addition to conventional morphological classification, we systematically search for the presence of irregular, asymmetric, features that are indicative of dynamical perturbations such as tidal interactions and galaxy-galaxy mergers.  As with the main morphological classification, three co-authors visually examine the \hst\ images based on four independent criteria. First, we divide the hosts into three groups depending on the presence of a tidal tail or shell.  Second, hosts showing clear signs of interactions are identified, as either a ``major merger'' for the hosts that exhibit interacting companions of comparable brightness, or ``minor merger'' for those containing an apparent companion of significantly lower brightness.  Third, by analogy with a similar classification scheme employed for luminous infrared galaxies (\citealt{haan_2011, stierwalt_2013}), we distinguish four merging stages: ``m1'' for hosts having a companion galaxy with no obvious signs of interaction; ``m2'' for hosts with signs of interaction, such as tidal tail or shell, but no major disturbances; ``m3'' for hosts with major disturbances but a companion galaxy that has not yet merged; ``and m4'' for hosts with two galaxies sharing a common envelope.  And  fourth, the hosts are split into three subgroups according to the presence of a double nucleus: ``s'' for hosts with a single nucleus; ``d'' for hosts with double nuclei residing in a single host; and ``dd'' for hosts with double nuclei in separate hosts. 

The classifications based on the merging properties are listed in Table~2.  
Example images of different subtypes are displayed in Figure~5, and Figure~6 
shows cutout images of hosts in different merging stages.  We find 38 objects 
that are accompanied by tidal tails (27 objects) or shells (11 objects). 
Based on signs of interactions, 38 objects are classified as merging galaxies, 
among them 18 major mergers and 20 minor mergers.  Intermediate and late 
merging stages (i.e. ``m2'', ``m3'', or ``m4'') can be assigned to 28 objects.
Note that this number of objects is slightly smaller than the values from 
other merging indicators (38) because the merging stage is defined only 
if the body of the companion galaxy is clearly seen in the science images. 
For example, a stellar stream or faint shell is not considered as a companion 
galaxy in this classification.  Finally, only three objects appear to have 
double nuclei in a single host (i.e. ``d''), while four objects have double 
nuclei in separate hosts (i.e. ``dd'').  We will discuss the physical 
connection between merging features and AGN properties in Section~{5.2}. 
The merging fraction as a function of different physical properties of AGN is 
summarized in Table~3.

\section{Discussion}

\subsection{Type~1 vs. Type~2}

Because our sample is identified from Swift-BAT hard X-ray data and targets were selected for observation based on HST scheduling criteria unrelated to any galaxy properties other than location on the sky, selection biases resulting from X-ray obscuration or other AGN properties are minimized. Therefore,
it should be possible to obtain a statistically meaningful comparison of the 
host properties of type~1 and type~2 AGN in this survey sample. We find that 
there is no significant difference in Hubble type between the hosts of the two 
types of AGN. Approximately $52, 41$, and $7\%$ of the type~1 targets and 
$44, 46,$ and 10\% of type~2 targets are hosted by early-type, late-type, and 
peculiar galaxies, respectively. However, for the low-luminosity AGN 
(\lbol $\le 10^{44.5}$ \lum), the early-type fraction 
for type~2 ($\sim48\%$) is marginally higher than that for type~1 ($\sim37\%$).
In other words, out of 35 early-type hosts for type~2 AGN, about half (18) 
contain low-luminosity AGN. On the contrary, out of 39 early-type hosts for 
type~1 AGN, only 18\%\ (7) contain low-luminosity AGN. 
However, this luminosity dependence is not clearly 
detected in late-type hosts. 
We also confirm this trend using the 
Kolmogorov-Smirnov test to examine the distribution of bolometric luminosity of 
early-type hosts, finding that the null hypothesis that the two subsamples 
(type~1 and type~2) are drawn from the same population can be rejected with a 
confidence level of 99.9\%. 

It is worthwhile to note that the mean \mbh\ for type~2 AGN
($10^{8.20\pm0.58}$\solmass) is marginally higher than that 
for type~1 AGN ($10^{8.04\pm0.78}$\solmass) by $\sim0.2$ dex. 
Therefore, the different levels of luminosity dependence of Hubble type 
between type~1 and type~2 AGN could originate from bias in the \mbh\ because 
galaxy morphology is a strong function of the stellar mass of the host. 
To test this 
hypothesis, we perform the same experiment using the subsample with \mbh\
ranging from $10^{7.5}$\solmass to $10^{9.5}$\solmass, where the \mbh\ 
distributions of type~1 and type~2 AGN are similar. We again find that 
there are different levels of luminosity dependence between type~1 and type~2 
AGN, indicating that such variation is not due to \mbh\ bias. 

The merging properties in the host can also provide insight into the 
physical connection between type~1 and type~2 AGN. In this study, we consider 
hosts in relatively late stages (i.e. ``m2'', ``m3'', or ``m4'' in merging 
stages) as merging hosts. The merging fractions for both types of AGN are 
nearly identical (16\%\ and 20\% for type~1 and type~2, respectively). 
This is somewhat inconsistent with previous studies of nearby obscured 
AGN, which claimed that obscured AGN are more preferentially found in merging 
galaxies compared with unobscured AGN (e.g., \citealt{donley_2018, 
ellison_2019}). This discrepancy could be due to the fact that IR-selected 
obscured AGN have higher merging fractions compared with optical
and X-ray selected AGN.   

We again divide the sample into two subgroups according 
to bolomteric luminosity. For the low-luminosity AGN (\lbol 
$\le 10^{44.5}$ \lum), the merging fraction for type~2 
($11^{+5.5}_{-4.4}$\%\footnote{Note that the error estimation in each 
bin is conducted using Jeffrey's confidence interval, which is known to be 
robust for a small sample size.}) 
is marginally higher than that for type~1 ($0^{+5.0}$\%). 
This luminosity dependence of the merging fraction is also seen in Figure~7. 
This trend is commonly detected in various merging features (e.g., the 
existence of a tidal tail or shell, major or 
minor merging, and the presence of double nuclei; Figure~7). More interestingly, this
discrepancy between type~1 and type~2 AGN disappears in the 
parameter space of Eddington ratio (Figure~8). Therefore, bolometric 
luminosity appears to be a main driver to determine the role merging
plays in type~1 compared with type~2 AGN.

In addition, the merging fraction of type~1 AGN peaks at a higher bolometric 
luminosity compared with that of type~2 AGN.
Taken at face value, this indicates either (1) the AGN triggering 
mechanism is not only dependent upon bolometric luminosity, but also 
differentiates somewhat between type~1 and type~2 AGN, or (2) the types are 
in different evolutionary stages, in the sense that type~2 precedes type~1.

\begin{figure*}[t!]
\centering
\includegraphics[width=1.0\textwidth]{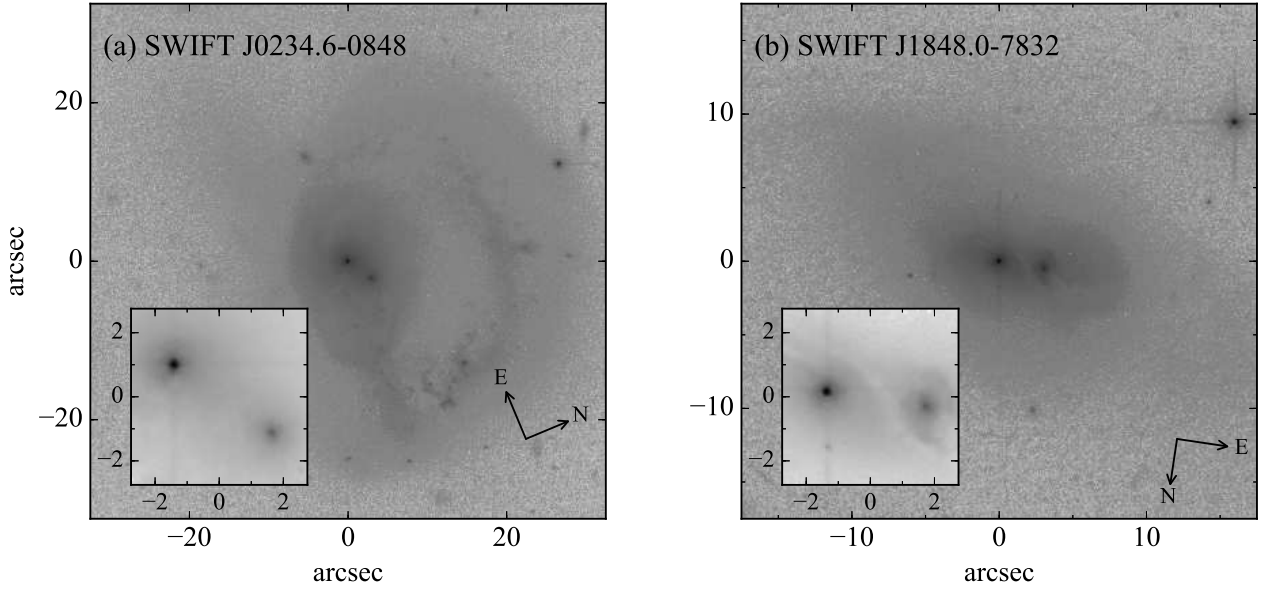}
\caption{
Double nuclei embedded in the host of (a) SWIFT J0234.6$-$0848 and (b) SWIFT J1848.0$-$7832.  A close-up view of the central region is shown in the inset.
}
\end{figure*}

\begin{figure}[t]
\centering
\includegraphics[width=0.45\textwidth]{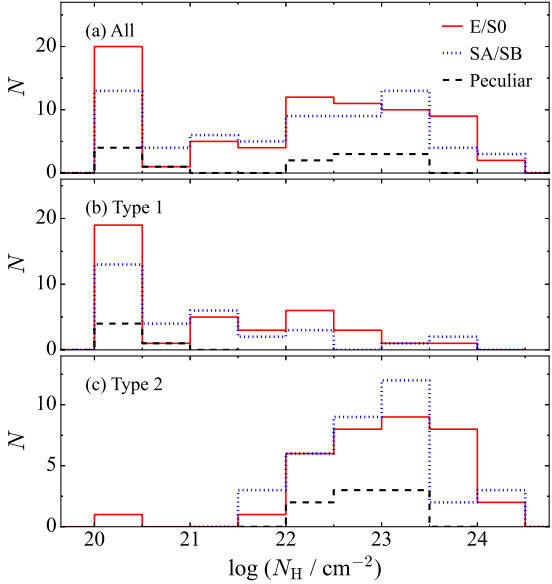}
\caption{
The distributions of the column density of neutral hydrogen ($N_{\rm H}$) derived from X-ray observations (Ricci et al. 2017) for different morphological classifications. The red, blue, and black histograms denote early-type galaxies (E/S0), late-type galaxies with spiral arms (SA/SB), and peculiar galaxies, respectively.
}
\end{figure}

\begin{figure}[t]
\centering
\includegraphics[width=0.45\textwidth]{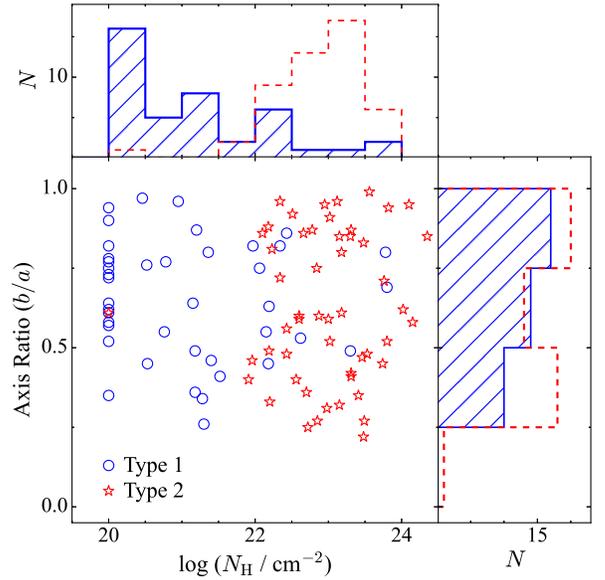}
\caption{
Distribution of host galaxt axis ratio ($b/a$) versus column density of neutral hydrogen ($N_{\rm H}$) derived from X-ray observations (Ricci et al. 2017) for type~1 AGN (blue stars and blue histograms) and type~2 (red circles and red histograms) AGN.
}
\end{figure}

\begin{figure*}[t!]
\centering
\includegraphics[width=1.0\textwidth]{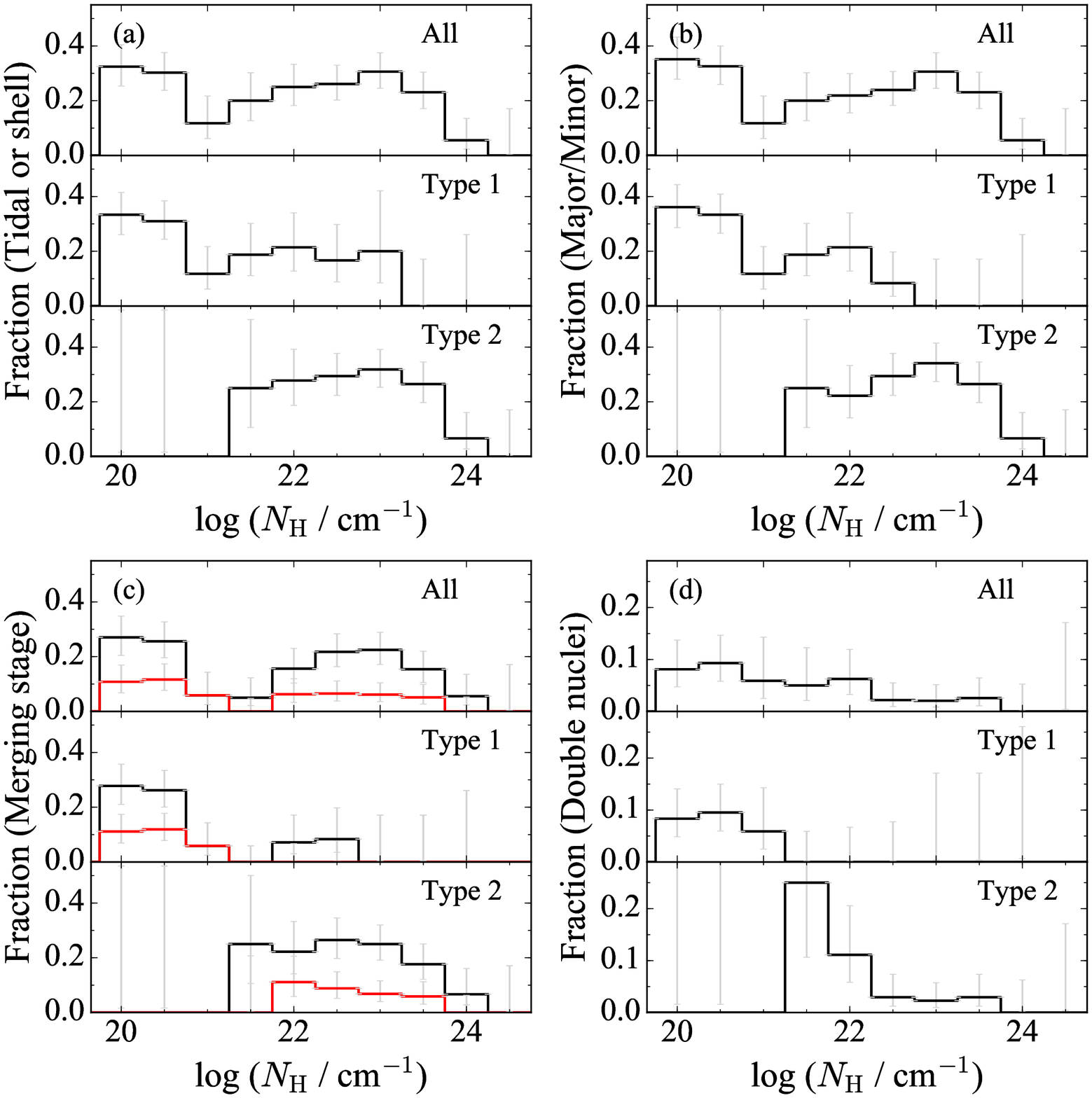}
\caption{
Dependence of the merging fraction on X-ray absorbing column density $N_{\rm H}$, for hosts with (a) a tidal tail or shell; (b) a major or minor merger; (c) tidal features that belong to merging stages ``m2'', ``m3'', or ``m4'' (black histograms) and those in the latest merging stages (``m4''; red histogram); (d) fraction of hosts having double nuclei (``d'' or ``dd'').  Each panel shows separate distributions for all AGN, type~1 AGN, and type~2 AGN. As described in the 
text, the fraction of galaxies found to contain double nuclei in this sample 
should be considered a lower limit, because the $I$-band imaging is not 
sufficient to identify some heavily obscured nuclei.
}
\end{figure*}

\subsection{Triggering Mechanism of AGN}
The relative importance of internal (secular) and external (merger-driven) 
triggering of AGN is still not fully understood. While internal
secular evolution in the host is thought to be responsible for enhancing the 
gas inflow toward a SMBH, for some AGN at least, galaxy merging is still 
regarded as a primary mechanism that triggers luminous AGN 
(e.g., \citealt{hong_2015, kim_2017, marian_2020}). The proximity of the 
target galaxies and deep \hst\ images 
with high spatial resolution make our dataset useful to identify 
merging features. In this study, we compare visually identified merging 
features and physical properties of AGN to investigate the role merging plays 
in triggering AGN in the Swift-BAT sample.

We use four criteria to identify the merging features; 
merging fraction and its trend are in very good agreement with one another 
(Figs. 7 and 8).  
While these visual classifications are somewhat subjective and cannot capture 
the full complexity of galaxy mergers, we use these classifications as a 
simple proxy for the merger stage. 
Out of 154 targets, 28 ($\sim18\%$) are classified as mergers (i.e., ``m2'', 
``m3'', or ``m4'' in the merging stage). This is in good 
agreement with the result ($\sim18\%$) derived using the SDSS images from a 
similar sample \citep{koss_2011}. \citet{zhao_2021} reached a similar
conclusion from \hst\ images of Palomar-Green QSOs at $z<0.5$; they found that 
only 17\%\ of the sample exhibits signs of interactions. 
However, our estimation of merging fraction is significantly lower than that 
for nearby luminous AGN (\citealt{hong_2015,gao_2020,marian_2020}).
More strikingly, the merging fraction for less luminous AGN in our sample 
(\lbol $\le 10^{44.5}$ \lum) is dramatically lower than that for more luminous 
AGN (\lbol $> 10^{44.5}$ \lum; $7.0^{+4.1}_{-2.6}$\%\ versus 
$25^{+4.6}_{-4.12}$\%).  
Previous studies have shown that the merging fraction can be as high as 
$\sim40-45$\%\ for relatively nearby luminous AGN (\lbol\ $> 10^{45}$ \lum). 
For example, \citet{hong_2015} claimed that 
the merger fraction of luminous AGN at $z<0.3$ is $\sim44$\%. 
\citet{marian_2020} analyzed ground-based optical images of luminous AGN
at $z<0.2$ and reached a similar conclusion (merging fraction of 
$\sim41$\%).
\citet{gao_2020} also showed that the merging fraction of less luminous 
AGN at $z<0.1$ (\lbol\ $<10^{45}$ \lum) is as low as $\sim20$\%\ and that for 
more luminous AGN at $z<0.6$ can be as high as $\sim40$\%. 
Using more distant AGN ($z\sim0.7$) but with moderate luminosity  
($10^{42.5} < $ \lbol\ $<10^{45.5}$ \lum), \citet{villforth_2014} found 
a merging fraction of $\sim17$\%. 
For our sample, the merging fraction for \lbol $\le 10^{45}$ \lum\ is 
$\sim17$\%, and that for \lbol\ $>10^{45}$ \lum\ is $\sim23$\%. Therefore,
our findings are in broad agreement with those of previous studies. However,
the merging fraction for luminous AGN appears to be systematically lower
than those from the literature. This discrepancy could be due to the fact
that our $z<0.1$ sample lacks highly luminous AGN (\lbol\ $>10^{45.5}$ \lum) 
that are better represented in more distant AGN samples.

As shown in Figure~7, the merging fraction dramatically 
increases as AGN luminosity increases. This trend is in good 
agreement with those of previous studies based on nearby and distant AGN 
(\citealt{treister_2012,kim_2017,ellison_2019}), suggesting that the luminous 
AGN are more likely to be triggered by merging. To the contrary, this 
tendency is not clearly observed in the comparison between the merging 
fraction and Eddington ratio (Figure~8). Taken at face value, the merging 
fraction also marginally increases as Eddington ratio increases. 
The merging fraction appears to peak
at around $\log$ \edd $\sim-1.5$. This trend is more prominent for type~1 AGN
although the significance of this finding should be further explored using 
larger sample sizes. 
It may also be related to the fact that NLS1, which tend
to have large Eddington ratio, are preferentially found in non-merger barred 
hosts, revealing that lower mass BH can be efficiently fueled at a high rate 
through secular processes (\citealt{crenshaw_2003, orban_2011, kim_2017}).
Nevertheless, this finding suggests that merging may not be 
necessary to enhance the Eddington ratio (see also \citealt{weigel_2018}).

Using near-infrared ($K-$band) images obtained with adaptive optics for 
Swift-BAT AGN, \citet{koss_2018} claimed that luminous obscured AGN have an 
excess of late-stage mergers with double nuclei with a close separation 
($d\le 3$kpc) compared to unobscured AGN and inactive galaxies. Due to the
high spatial resolution of the \hst\ images, our dataset is suitable to test 
this finding independently, even though the obscuration tends to be more 
severe in 
$I-$ band compared with $K-$band. In this study, we find only three objects 
(SWIFT~J0234.6$-$0848, SWIFT~J1845.4+7211, and 
SWIFT~J1848.0$-$7832; Figure~9) that have double nuclei in a single merged host 
(i.e., ``d'' in the number of nuclei). The separations between the two nuclei
are 3\farcs7 (2.6 kpc), 1\farcs6 (1.5 kpc), and 3\farcs1 (4.5 kpc) for 
SWIFT~J0234.6$-$0848, SWIFT~J1845.4+7211, and SWIFT~J1848.0$-$7832, 
respectively. Thus, two objects from our survey (SWIFT~J0234.6$-$0848 and 
SWIFT~J1845.4+7211) satisfy 
the criteria ($d\le 3$ kpc) for close-separation double nuclei from \citet{koss_2018}. For comparison, two objects (SWIFT~J0804.6+1045 and SWIFT~J1845.4+7211) classified as 
close nuclei in \citet{koss_2018} are 
also observed in this study. While SWIFT~J1845.4+7211 is classified as ``d'' 
based on the nuclear structure visible in the HST F814W data, SWIFT~J0804.6+1045 appears to have a single 
nucleus in the F814W image. This indicates that $I$-band images will 
miss some double nuclei that can be identified in longer-wavelength imaging at 
comparable resolution, as would be expected for gas-rich mergers with highly 
obscured central regions.

Based on ground-based imaging surveys of nearby \swift-BAT AGN,
several studies have independently examined the merging fraction using different
methods \citep[e.g.,][]{koss_2010, koss_2012, cotini_2013}. For example, 
\citet{koss_2010} found that the fractions of disturbed host galaxies and
of close pairs within 30 kpc in nearby sources ($z<0.05$) are $\sim18$\% and 
$\sim24$\%, respectively. The merging fraction coincides with our finding.
The fraction of close pairs from this study ($\sim33$\%) is marginally higher 
than that from \citet{koss_2010}, although we do not impose any constraint 
on the distance to the companion, and the imaging quality of the two studies in terms of the 
depth and spatial resolution is not identical. 
\citet{cotini_2013} analyzed ground-based images of nearby sources 
($z<0.03$) to identify interacting systems through quantitative methods
\citep{conselice_2003, lotz_2004}. They found that the merging fraction 
is $20^{+7}_{-5}$\%, again in good agreement with this study.  
Finally, \citet{koss_2012} used optical spectroscopy and X-ray data to 
identify dual AGNs from the nearby \swift-BAT AGN ($z<0.05$), and found that
$\sim10$\% of the sample contains dual AGN. They also found that the fraction 
of dual AGN increases with increasing AGN luminosity, supporting the idea
that merging plays an important role in luminous AGN.

\subsection{Connection between the Obscuration and Merging Properties}
As discussed in Section~{5.1} and claimed in previous studies, there are 
marginal differences in the host properties between type~1 and type~2 AGN. 
Observational studies have found that significant obscuration can originate 
from dense material on larger scales in the host galaxy (e.g., 
\citealt{simcoe_1997,malkan_1998,malizia_2020}). In addition, theoretical and 
observational studies have suggested that the central BH can be heavily 
obscured during mergers (\citealt{sanders_1988, blecha_2018}). The hard X-ray 
selection makes the Swift-BAT sample particularly useful for exploring whether 
there is any connection between the degree of obscuration and the host galaxy 
morphological type or inclination.
Here, we compare the 
hydrogen column density ($N_{\rm H}$) as a proxy of obscuration with the host 
properties of our sample.  

Figure~10 shows the distributions of column density for different morphologies.
Based on the Kolmogorov–Smirnov test, there is no significant 
difference in $N_{\rm H}$ distributions between Hubble types. More surprisingly,
the majority of highly obscured AGN ($\log N_{\rm H} > 10^{23}$ cm$^{-1}$) are
hosted by early-type galaxies. 
In addition, we compare the axis ratio with the hydrogen column density 
(Figure~11), and find there is no casual correlation between the two, and 
there is little difference in the axis ratio distributions between type~1 and 
type~2 AGN. These findings might indicate that the obscuration material is 
not physically related to the host galaxy, but rather to the nuclear torus.

We also compare the merging properties with $N_{\rm H}$ (Figure~12).
As type~1 AGN tend to have a smaller $N_{\rm H}$ than type~2 by nature, it is 
not adequate to directly compare trends of type~1 and those of type~2 AGN. 
However, it is intriguing that the merging fractions appear to peak at 
$\log N_{\rm H} \sim 10^{20}$ cm$^{-1}$ and $\log N_{\rm H} \sim 10^{22-23}$ 
cm$^{-1}$. However, it is difficult to understand the origin of this 
bimodality; it might imply that the obscuration is somehow related to the 
various types of merging properties (e.g., gas contents, orbital motion, 
mass ratio, and merging stages). Contrary to our findings, previous 
observational studies 
claimed that obscured AGN, as opposed to unobscured AGN, are more likely 
to be hosted by galaxies that exhibit merging features (e.g., 
\citealt{urrutia_2008,kocevski_2015}). However, because previous studies 
primarily relied on red AGN selected from optical and 
IR bands (\citealt{glikman_2015}) or distant X-ray selected AGN ($z>1$; 
\citealt{kocevski_2015}), the samples could be biased toward either luminous 
AGN or high$-z$ environments, where merging plays a more important role
in BH growth than at the present epoch. 

\section{Summary}
Through an intensive imaging survey of \swift-BAT AGN at $z<0.1$ using the \hst,
we obtained coherent, deep $I-$band images for 154 targets. We conducted a 
visual classification of the host galaxies based on 
Hubble type and merging features. We also assembled AGN properties (BH mass, 
bolometric luminosity, Eddington ratio, and neutral hydrogen column density) 
of the sample from the literature, and compare them with the 
morphological properties of the hosts. From this program, we obtain
the following results:

\begin{itemize}

\item Out of 154 targets, 74, 67, and 13 hosts 
are classified as early-type (E/S0), late-type (SA/SB) and peculiar galaxies, 
respectively. 

\item Using merger indicators including tidal tails and shells, 
we find 18\%$-$25\% of the 
sample is hosted by the merging galaxies, which is broadly consistent with 
the findings of previous studies. However, the merging fraction substantially 
is larger for high-luminosity AGN (\lbol\ $> 10^{44.5}$) compared with less 
luminous AGN \lbol\ $\le 10^{44.5}$ (25\%\ versus 7\%, respectively). It 
suggests that the merging plays a more important role 
in triggering luminous AGN in the Swift-BAT sample. 

\item To test AGN unification models, we compare the host morphologies of
Type~1 and 2 AGN. We find there is a negligible difference in terms of
Hubble type and merging fraction between these hosts. The merging
fraction for Type~2 AGN, however, peaks at a marginally lower AGN luminosity 
than that of Type~1 AGN, implying that the triggering mechanism and/or 
evolutionary stage for Type~1 and Type~2 may differ somewhat. 

\item Finally, we compare the neutral hydrogen column density and axis 
ratio to host properties to investigate the origin of the obscuration. 
Although we did not identify any clear trend as a result of this comparison, 
we find that merging fractions appear to reach their
maximum at $\log N_{\rm H} \sim 10^{20}$ cm$^{-1}$ and $\log N_{\rm H} 
\sim 10^{22-23}$. Although this bimodality has yet to be 
confirmed with a larger sample size, it could suggest that the 
physical origin of obscuration is complex.  
\end{itemize}

The dataset presented in this paper will have broad applications for 
addressing other questions related to  the physical connection between SMBHs 
and ther host galaxies. 
For example, we are currently performing two-dimensional imaging decomposition, 
and examining the \mlb\ relation for type~1 AGN. Our sample can also be 
combined with existing archival \hst\ data of other \swift-BAT AGN, 
which will enable us to investigate the host galaxy statistical properties 
with a larger sample size. Finally, our extensive imaging survey can be used 
to carry out detailed host-galaxy studies of individual targets of interest 
among the nearby AGN population \citep[e.g.,][]{maksym_2020}.

\acknowledgments
We are grateful to the anonymous referee for the constructive comments. 
We are grateful to STScI for supporting the gap-filler snapshot programs, 
and we thank Amber Armstrong and Norman Grogin for their assistance in 
preparing the Phase II program for this survey.
LCH was supported by the National Science Foundation of China (11721303, 
11991052, 12011540375) and the National Key R\&D Program of China 
(2016YFA0400702). This work was supported by a National Research Foundation of 
Korea (NRF) grant (No.\  2020R1A2C4001753) funded by the Korean government 
(MSIT) and under the framework of international cooperation program managed 
by the National Research Foundation of Korea (NRF-2020K2A9A2A06026245).  

\bibliography{host}

\figsetstart
\figsetnum{13}
\figsettitle{Images}

\figsetgrpstart
\figsetgrpnum{13.1}
\figsetgrptitle{$HST$ F814W-band Images of SWIFT J0001.0$-$0708, SWIFT J0001.6$-$7701, SWIFT J0003.3+2737, and SWIFT J0005.0+7021.}
\figsetplot{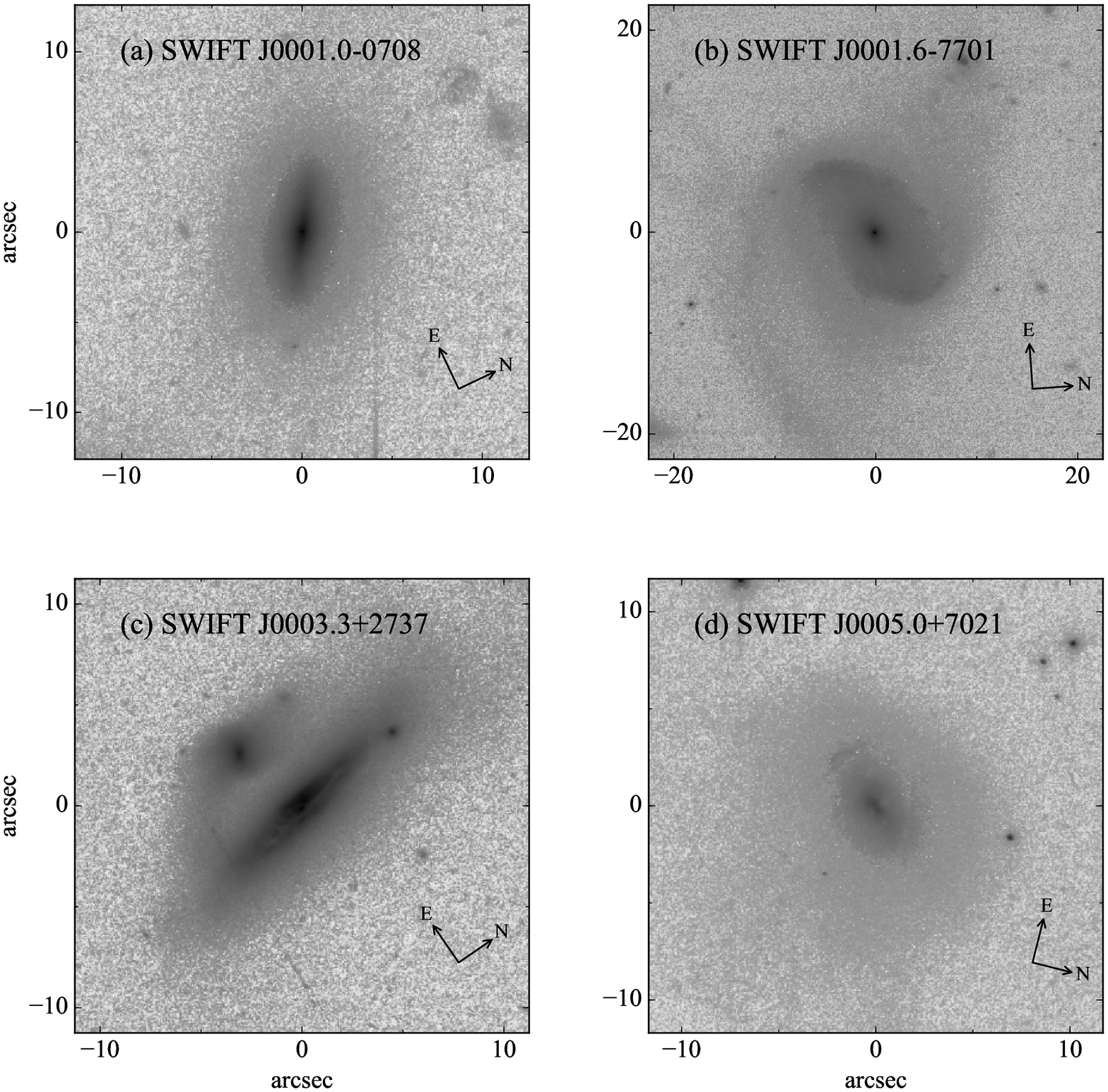}
\figsetgrpnote{\hst\ F814W-band images of the target galaxies.  Images for the entire sample are available in the online figure set.}
\figsetgrpend

\figsetend

\begin{figure}
\figurenum{13}
\plotone{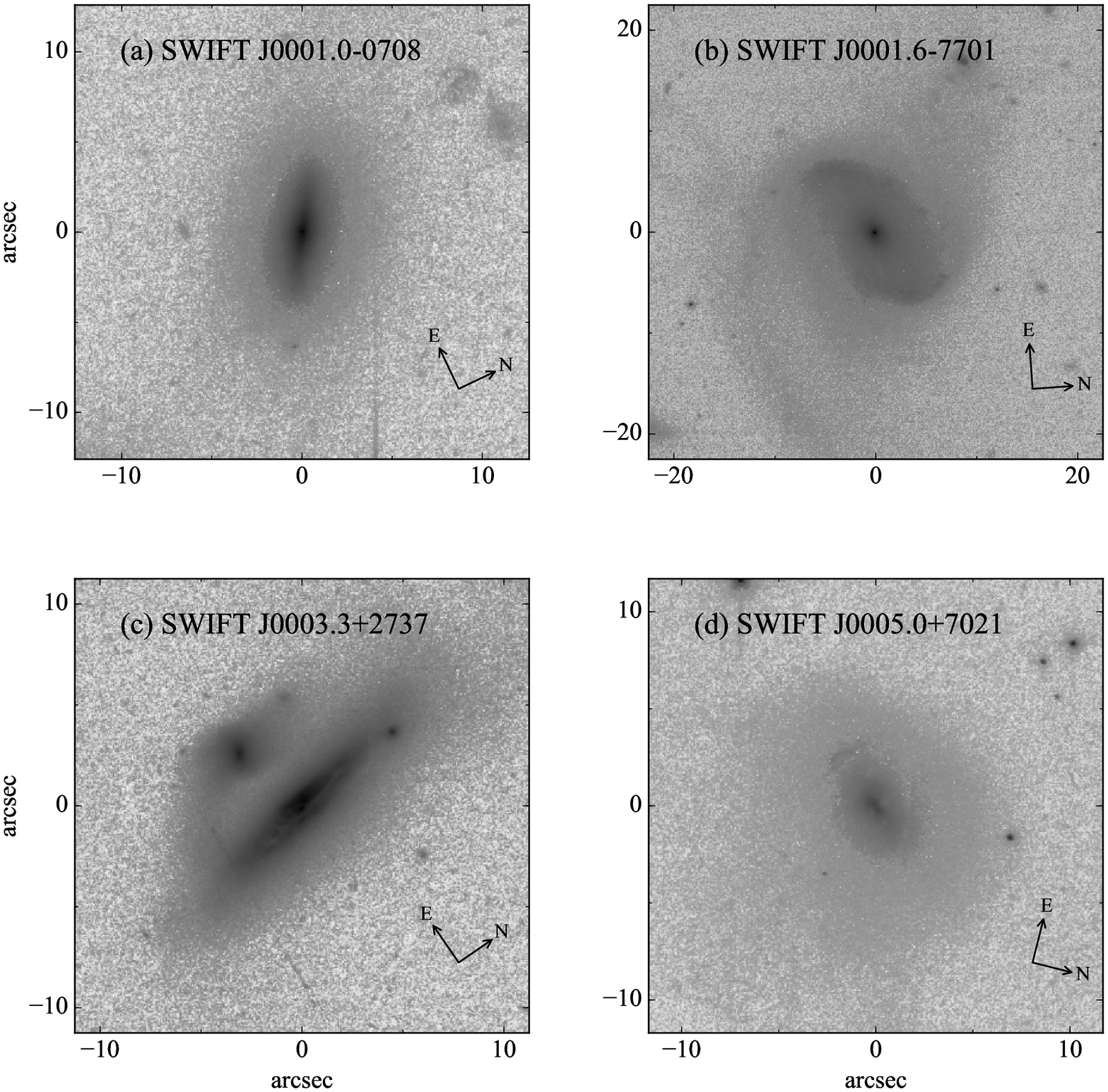}
\caption{\hst\ F814W-band images of the target galaxies.  Images for the entire sample are available in the online figure set.}
\end{figure}

\startlongtable
\begin{longrotatetable}
\begin{deluxetable}{llrrrccccccccc}
\tablecolumns{14}
\tabletypesize{\footnotesize}
\tablewidth{0pc}
\tablecaption{The sample}
\tablehead{
\colhead{\h Source Name} &
\colhead{\h Alt. Name} &
\colhead{\h R.A.} &
\colhead{\h Dec.} &
\colhead{\h $A_V$} &
\colhead{\h $z$} &
\colhead{\h Type} &
\colhead{\h Obs. Date} &
\colhead{\h log $L_{\rm bol}$} &
\colhead{\h log $M_{\rm BH}$} &
\colhead{\h M} &
\colhead{\h log $\lambda$} &
\colhead{\h $b/a$} &
\colhead{\h log $N_{\rm H}$} \\
\colhead{\h } &
\colhead{\h } &
\colhead{\h (deg.)} &
\colhead{\h (deg.)} &
\colhead{\h (mag)} &
\colhead{\h } &
\colhead{\h } &
\colhead{\h } &
\colhead{\h (erg s$^{-1}$)} &
\colhead{\h ($M_{\odot}$)} &
\colhead{\h } &
\colhead{\h } &
\colhead{\h } &
\colhead{\h (cm$^{-2}$)} \\
\colhead{\h (1)} &
\colhead{\h (2)} &
\colhead{\h (3)} &
\colhead{\h (4)} &
\colhead{\h (5)} &
\colhead{\h (6)} &
\colhead{\h (7)} &
\colhead{\h (8)} &
\colhead{\h (9)} &
\colhead{\h (10)} &
\colhead{\h (11)} &
\colhead{\h (12)} &
\colhead{\h (13)} &
\colhead{\h (14)}
}
\startdata
SWIFT J0001.0$-$0708&\h &\h   0.2032&\h  -7.1532&\h 0.087&\h 0.0375&\h 1&\h 2018-12-05&\h 44.50&\h  7.12&\h H$\alpha$&\h  -0.72&\h 0.63&\h $22.19^{+0.07}_{-0.08}$ \\
SWIFT J0001.6$-$7701&\h Fairall 1203&\h   0.4419&\h -76.9540&\h 0.160&\h 0.0585&\h 1&\h 2018-12-23&\h 44.82&\h  8.75&\h $M_*$&\h  -2.03&\h 0.90&\h $20.00^{+0.00}_{-0.00}$ \\
SWIFT J0003.3+2737&\h &\h   0.8643&\h  27.6548&\h 0.164&\h 0.0397&\h 2&\h 2019-01-22&\h 44.57&\h  7.85&\h $\sigma_*$&\h  -1.38&\h 0.37&\h $22.86^{+0.12}_{-0.08}$ \\
SWIFT J0005.0+7021&\h &\h   1.0082&\h  70.3217&\h 2.313&\h 0.0964&\h 1&\h 2018-12-02&\h 45.33&\h  8.20&\h $M_*$&\h  -0.97&\h 0.87&\h $22.61^{+0.07}_{-0.05}$ \\
SWIFT J0006.2+2012&\h Mrk 335&\h   1.5813&\h  20.2029&\h 0.096&\h 0.0259&\h 1&\h 2018-05-20&\h 44.30&\h  7.34&\h H$\alpha$&\h  -1.14&\h 0.87&\h $20.48^{+0.03}_{-0.22}$ \\
SWIFT J0036.3+4540&\h &\h   9.0874&\h  45.6650&\h 0.181&\h 0.0476&\h 1&\h 2019-01-12&\h 44.80&\h  8.54&\h $M_*$&\h  -1.84&\h 0.52&\h $20.00^{+0.00}_{-0.00}$ \\
SWIFT J0100.9$-$4750&\h ESO 195-IG 021 NED03&\h  15.1457&\h -47.8676&\h 0.037&\h 0.0484&\h 2&\h 2018-03-31&\h 44.84&\h  8.42&\h $\sigma_*$&\h  -1.68&\h 0.39&\h $22.57^{+0.03}_{-0.04}$ \\
SWIFT J0123.9$-$5846&\h Fairall 9&\h  20.9408&\h -58.8057&\h 0.071&\h 0.0460&\h 1&\h 2018-12-25&\h 45.32&\h  8.38&\h H$\alpha$&\h  -1.16&\h 0.85&\h $20.00^{+0.00}_{-0.00}$ \\
SWIFT J0128.4+1631&\h CGCG 459-058&\h  22.1017&\h  16.4593&\h 0.199&\h 0.0387&\h 1&\h 2019-01-22&\h 44.57&\h  6.51&\h H$\alpha$&\h  -0.04&\h 0.49&\h $23.30^{+0.18}_{-0.04}$ \\
SWIFT J0134.1$-$3625&\h NGC 612&\h  23.4906&\h -36.4933&\h 0.055&\h 0.0298&\h 2&\h 2018-12-25&\h 45.23&\h  8.99&\h $\sigma_*$&\h  -1.86&\h 0.86&\h $23.97^{+0.11}_{-0.07}$ \\
SWIFT J0157.2+4715&\h &\h  29.2956&\h  47.2666&\h 0.409&\h 0.0478&\h 1&\h 2019-01-13&\h 44.84&\h  7.85&\h H$\alpha$&\h  -1.11&\h 0.82&\h $20.00^{+0.00}_{-0.00}$ \\
SWIFT J0202.4+6824A&\h &\h  30.5720&\h  68.3620&\h 3.094&\h 0.0119&\h 2&\h 2019-01-20&\h 43.38&\h  8.94&\h $M_*$&\h  -3.66&\h 0.81&\h $22.23^{+0.35}_{-0.45}$ \\
SWIFT J0202.4+6824B&\h &\h  30.3850&\h  68.4061&\h 3.085&\h 0.0152&\h 2&\h 2019-01-20&\h 43.50&\h  9.04&\h $M_*$&\h  -3.64&\h 0.86&\h $22.10^{+0.13}_{-0.69}$ \\
SWIFT J0206.2$-$0019&\h Mrk 1018&\h  31.5666&\h  -0.2914&\h 0.075&\h 0.0430&\h 1&\h 2019-01-11&\h 45.01&\h  8.00&\h H$\alpha$&\h  -1.09&\h 0.64&\h $20.00^{+0.00}_{-0.00}$ \\
SWIFT J0234.6$-$0848&\h NGC 985&\h  38.6574&\h  -8.7876&\h 0.090&\h 0.0430&\h 1&\h 2019-01-16&\h 45.02&\h  8.43&\h H$\alpha$&\h  -1.51&\h 0.93&\h $20.92^{+0.09}_{-0.06}$ \\
SWIFT J0243.9+5324&\h 2MFGC 02171&\h  41.0125&\h  53.4745&\h 1.040&\h 0.0364&\h 2&\h 2019-01-12&\h 44.29&\h  7.83&\h $M_*$&\h  -1.64&\h 0.25&\h $22.72^{+0.17}_{-0.12}$ \\
SWIFT J0333.3+3720&\h &\h  53.3282&\h  37.3030&\h 1.465&\h 0.0547&\h 1&\h 2019-01-11&\h 45.14&\h  8.15&\h H$\alpha$&\h  -1.11&\h 0.80&\h $21.36^{+0.16}_{-0.20}$ \\
SWIFT J0347.0$-$3027&\h &\h  56.7729&\h -30.3972&\h 0.030&\h 0.0950&\h 1&\h 2018-12-21&\h 45.18&\h  7.76&\h $M_*$&\h  -0.68&\h 0.72&\h $20.00^{+0.00}_{-0.00}$ \\
SWIFT J0356.9$-$4041&\h &\h  59.2356&\h -40.6960&\h 0.023&\h 0.0747&\h 2&\h 2018-12-19&\h 45.22&\h  8.74&\h $M_*$&\h  -1.62&\h 0.86&\h $22.66^{+0.03}_{-0.04}$ \\
SWIFT J0405.3$-$3707&\h ESO 359- G 019&\h  61.2570&\h -37.1875&\h 0.016&\h 0.0552&\h 1&\h 2019-01-13&\h 44.77&\h  8.18&\h $M_*$&\h  -1.51&\h 0.61&\h $20.00^{+0.00}_{-0.00}$ \\
SWIFT J0429.6$-$2114&\h &\h  67.4095&\h -21.1622&\h 0.067&\h 0.0700&\h 1&\h 2018-11-26&\h 45.00&\h  8.72&\h H$\alpha$&\h  -1.82&\h 0.70&\h $20.00^{+0.00}_{-0.00}$ \\
SWIFT J0443.9+2856&\h &\h  70.9450&\h  28.9718&\h 2.028&\h 0.0217&\h 1&\h 2018-11-13&\h 44.48&\h  8.08&\h $M_*$&\h  -1.70&\h 0.82&\h $22.34^{+0.03}_{-0.03}$ \\
SWIFT J0446.4+1828&\h UGC 3157&\h  71.6240&\h  18.4609&\h 1.646&\h 0.0158&\h 1&\h 2019-01-21&\h 43.91&\h  6.69&\h H$\alpha$&\h  -0.88&\h 0.80&\h $23.78^{+0.21}_{-0.19}$ \\
SWIFT J0456.3$-$7532&\h ESO 033- G 002&\h  73.9957&\h -75.5412&\h 0.400&\h 0.0181&\h 2&\h 2018-12-14&\h 44.13&\h  8.44&\h $M_*$&\h  -2.41&\h 0.92&\h $22.51^{+0.40}_{-0.31}$ \\
SWIFT J0504.6$-$7345&\h &\h  76.1425&\h -73.8242&\h 0.335&\h 0.0451&\h 2&\h 2020-10-02&\h 44.52&\h  8.58&\h $M_*$&\h  -2.16&\h 0.40&\h $21.91^{+0.14}_{-0.14}$ \\
SWIFT J0510.7+1629&\h IRAS 05078+1626&\h  77.6896&\h  16.4989&\h 0.820&\h 0.0173&\h 1&\h 2019-01-19&\h 44.72&\h  7.64&\h H$\alpha$&\h  -1.02&\h 0.66&\h $21.08^{+0.07}_{-0.10}$ \\
SWIFT J0516.2$-$0009&\h Ark 120&\h  79.0476&\h  -0.1498&\h 0.349&\h 0.0325&\h 1&\h 2018-11-28&\h 45.11&\h  8.68&\h H$\alpha$&\h  -1.67&\h 0.82&\h $20.00^{+0.00}_{-0.00}$ \\
SWIFT J0526.2$-$2118&\h ESO 553- G 043&\h  81.6135&\h -21.2866&\h 0.116&\h 0.0277&\h 2&\h 2019-01-22&\h 44.30&\h  8.05&\h $\sigma_*$&\h  -1.85&\h 0.85&\h $23.30^{+0.07}_{-0.07}$ \\
SWIFT J0528.1$-$3933&\h &\h  82.0086&\h -39.5791&\h 0.073&\h 0.0369&\h 2&\h 2020-05-31&\h 44.37&\h  8.20&\h $M_*$&\h  -1.93&\h 0.49&\h $21.68^{+0.22}_{-0.28}$ \\
SWIFT J0533.9$-$1318&\h &\h  83.4589&\h -13.3553&\h 0.371&\h 0.0290&\h 2&\h 2019-10-23&\h 44.52&\h  8.51&\h $M_*$&\h  -2.09&\h 0.71&\h $23.76^{+0.19}_{-0.33}$ \\
SWIFT J0544.4$-$4328&\h &\h  86.0000&\h -43.4240&\h 0.132&\h 0.0446&\h 2&\h 2021-04-07&\h 44.45&\h  7.40&\h $M_*$&\h  -1.05&\h 0.46&\h $22.81^{+0.14}_{-0.19}$ \\
SWIFT J0552.5+5929&\h &\h  88.1169&\h  59.4756&\h 0.511&\h 0.0585&\h 1&\h 2018-11-19&\h 44.96&\h  8.35&\h $M_*$&\h  -1.49&\h 0.82&\h $20.00^{+0.00}_{-0.00}$ \\
SWIFT J0623.9$-$6058&\h ESO 121-IG 028&\h  95.9400&\h -60.9790&\h 0.158&\h 0.0405&\h 2&\h 2018-12-14&\h 44.98&\h  9.04&\h $M_*$&\h  -2.16&\h 0.87&\h $23.31^{+0.05}_{-0.04}$ \\
SWIFT J0641.3+3257&\h &\h 100.3252&\h  32.8254&\h 0.414&\h 0.0486&\h 2&\h 2018-12-22&\h 45.15&\h  8.12&\h $\sigma_*$&\h  -1.07&\h 0.96&\h $23.12^{+0.05}_{-0.06}$ \\
SWIFT J0645.9+5303&\h &\h 101.6108&\h  53.0758&\h 0.207&\h 0.0359&\h 2&\h 2020-04-14&\h 44.41&\h  8.06&\h $M_*$&\h  -1.75&\h 0.96&\h $22.34^{+0.16}_{-0.15}$ \\
SWIFT J0707.1+6433&\h &\h 106.8047&\h  64.5997&\h 0.104&\h 0.0797&\h 1&\h 2019-01-16&\h 45.10&\h  8.56&\h $M_*$&\h  -1.56&\h 0.90&\h $20.00^{+0.00}_{-0.00}$ \\
SWIFT J0709.0$-$4642&\h &\h 107.1803&\h -46.7137&\h 0.298&\h 0.0468&\h 2&\h 2018-12-22&\h 44.74&\h  8.25&\h $\sigma_*$&\h  -1.61&\h 0.66&\h $22.40^{+0.16}_{-0.12}$ \\
SWIFT J0736.9+5846&\h Mrk 9&\h 114.2374&\h  58.7704&\h 0.160&\h 0.0399&\h 1&\h 2019-01-11&\h 44.40&\h  7.84&\h H$\alpha$&\h  -1.54&\h 0.97&\h $20.46^{+0.28}_{-0.31}$ \\
SWIFT J0743.0+6513&\h Mrk 78&\h 115.6739&\h  65.1771&\h 0.097&\h 0.0366&\h 2&\h 2019-01-06&\h 44.49&\h  8.27&\h $M_*$&\h  -1.88&\h 0.58&\h $24.15^{+0.11}_{-0.13}$ \\
SWIFT J0743.3$-$2546&\h &\h 115.8114&\h -25.7639&\h 1.969&\h 0.0238&\h 1&\h 2018-12-17&\h 44.30&\h  7.39&\h H$\alpha$&\h  -1.19&\h 0.96&\h $20.95^{+0.25}_{-1.65}$ \\
SWIFT J0747.5+6057&\h Mrk 10&\h 116.8714&\h  60.9335&\h 0.128&\h 0.0294&\h 1&\h 2019-01-10&\h 44.36&\h  7.54&\h H$\alpha$&\h  -1.28&\h 0.45&\h $20.53^{+0.22}_{-0.12}$ \\
SWIFT J0747.6$-$7326&\h &\h 116.9098&\h -73.4314&\h 1.128&\h 0.0359&\h 2&\h 2019-12-27&\h 44.61&\h  8.84&\h $M_*$&\h  -2.33&\h 0.99&\h $23.56^{+0.19}_{-0.16}$ \\
SWIFT J0753.1+4559&\h B3 0749+460A&\h 118.1842&\h  45.9493&\h 0.226&\h 0.0516&\h 1&\h 2019-01-08&\h 44.83&\h  8.38&\h H$\alpha$&\h  -1.65&\h 0.94&\h $20.00^{+0.00}_{-0.00}$ \\
SWIFT J0756.3$-$4137&\h &\h 119.0817&\h -41.6284&\h 2.116&\h 0.0210&\h 2&\h 2019-01-16&\h 43.90&\h  6.94&\h H$\alpha$&\h  -1.14&\h 0.56&\h $21.77^{+0.07}_{-0.07}$ \\
SWIFT J0759.8$-$3844&\h &\h 119.9242&\h -38.7322&\h 2.225&\h 0.0402&\h 1&\h 2019-02-03&\h 45.15&\h  8.52&\h H$\alpha$&\h  -1.47&\h 0.73&\h $20.00^{+0.00}_{-0.00}$ \\
SWIFT J0800.1+2638&\h IC 0486&\h 120.0874&\h  26.6135&\h 0.111&\h 0.0266&\h 1&\h 2018-12-29&\h 44.59&\h  7.06&\h H$\alpha$&\h  -0.57&\h 0.75&\h $22.06^{+0.12}_{-0.11}$ \\
SWIFT J0801.9$-$4946&\h &\h 120.4915&\h -49.7784&\h 0.709&\h 0.0405&\h 1&\h 2020-05-11&\h 44.80&\h  9.23&\h $M_*$&\h  -2.53&\h 0.36&\h $21.18^{+0.14}_{-0.18}$ \\
SWIFT J0804.6+1045&\h MCG +02-21-013&\h 121.1933&\h  10.7767&\h 0.073&\h 0.0349&\h 2&\h 2018-12-28&\h 44.55&\h  8.21&\h $\sigma_*$&\h  -1.76&\h 0.41&\h $23.02^{+0.15}_{-0.22}$ \\
SWIFT J0805.1$-$0110&\h &\h 121.2208&\h  -1.1466&\h 0.071&\h 0.0915&\h 2&\h 2021-03-26&\h 45.39&\h  8.45&\h $\sigma_*$&\h  -1.16&\h 0.47&\h $23.46^{+0.16}_{-0.31}$ \\
SWIFT J0807.9+3859&\h Mrk 622&\h 121.9210&\h  39.0042&\h 0.138&\h 0.0232&\h 2&\h 2018-05-25&\h 44.32&\h  6.74&\h H$\alpha$&\h  -0.52&\h 0.95&\h $24.10^{+1.90}_{-0.43}$ \\
SWIFT J0819.2$-$2259&\h &\h 124.7410&\h -22.8770&\h 0.335&\h 0.0346&\h 1&\h 2020-05-03&\h 44.39&\h  8.31&\h $M_*$&\h  -2.02&\h 0.75&\h $20.00^{+0.00}_{-0.00}$ \\
SWIFT J0823.4$-$0457&\h Fairall 272&\h 125.7546&\h  -4.9349&\h 0.125&\h 0.0222&\h 2&\h 2018-04-08&\h 44.66&\h  8.14&\h $\sigma_*$&\h  -1.58&\h 0.48&\h $23.53^{+0.05}_{-0.04}$ \\
SWIFT J0840.2+2947&\h 4C +29.30&\h 130.0099&\h  29.8174&\h 0.153&\h 0.0647&\h 2&\h 2019-01-21&\h 45.28&\h  8.77&\h $\sigma_*$&\h  -1.59&\h 0.85&\h $23.80^{+0.11}_{-0.10}$ \\
SWIFT J0855.6+6425&\h MCG +11-11-032&\h 133.8023&\h  64.3959&\h 0.279&\h 0.0362&\h 2&\h 2019-01-05&\h 44.66&\h  8.35&\h $\sigma_*$&\h  -1.79&\h 0.42&\h $23.31^{+0.13}_{-0.13}$ \\
SWIFT J0902.7$-$4814&\h &\h 135.6555&\h -48.2261&\h 4.222&\h 0.0391&\h 1&\h 2020-07-31&\h 44.90&\h  5.54&\h $M_*$&\h   1.26&\h 0.77&\h $22.15^{+0.13}_{-0.22}$ \\
SWIFT J0923.7+2255&\h MCG +04-22-042&\h 140.9292&\h  22.9090&\h 0.120&\h 0.0333&\h 1&\h 2019-01-11&\h 44.87&\h  7.54&\h H$\alpha$&\h  -0.77&\h 0.57&\h $20.00^{+0.00}_{-0.00}$ \\
SWIFT J0924.2$-$3141&\h &\h 140.9739&\h -31.6919&\h 0.408&\h 0.0426&\h 2&\h 2020-04-27&\h 45.00&\h  8.05&\h $M_*$&\h  -1.15&\h 0.75&\h $24.42^{+0.09}_{-0.07}$ \\
SWIFT J0925.2$-$8423&\h &\h 141.5735&\h -84.3593&\h 0.540&\h 0.0629&\h 1&\h 2018-12-28&\h 44.85&\h  7.81&\h $M_*$&\h  -1.06&\h 0.78&\h $22.40^{+0.08}_{-0.12}$ \\
SWIFT J0936.2$-$6553&\h &\h 144.0260&\h -65.8093&\h 0.640&\h 0.0393&\h 2&\h 2018-12-25&\h 44.50&\h  7.97&\h $M_*$&\h  -1.57&\h 0.59&\h $22.60^{+0.35}_{-2.60}$ \\
SWIFT J0942.2+2344&\h CGCG 122-055&\h 145.5200&\h  23.6853&\h 0.069&\h 0.0217&\h 1&\h 2019-01-19&\h 43.98&\h  7.35&\h H$\alpha$&\h  -1.47&\h 0.77&\h $20.78^{+0.33}_{-1.22}$ \\
SWIFT J0947.6$-$3057&\h MCG -05-23-016&\h 146.9173&\h -30.9489&\h 0.295&\h 0.0083&\h 1&\h 2019-01-23&\h 44.40&\h  6.62&\h H$\alpha$&\h  -0.32&\h 0.45&\h $22.18^{+0.03}_{-0.04}$ \\
SWIFT J1020.5$-$0237B&\h &\h 154.9941&\h  -2.5767&\h 0.112&\h 0.0595&\h 1&\h 2019-01-19&\h 44.60&\h  8.58&\h H$\beta$&\h  -2.08&\h 0.94&\h $20.00^{+0.00}_{-0.00}$ \\
SWIFT J1021.7$-$0327&\h ARK 241&\h 155.4177&\h  -3.4539&\h 0.123&\h 0.0410&\h 1&\h 2019-01-17&\h 44.72&\h  8.15&\h $M_*$&\h  -1.53&\h 0.89&\h $20.00^{+0.00}_{-0.00}$ \\
SWIFT J1029.8$-$3821&\h ESO 317- G 038&\h 157.4400&\h -38.3486&\h 0.199&\h 0.0151&\h 2&\h 2021-05-11&\h 43.65&\h  7.41&\h $\sigma_*$&\h  -1.86&\h 0.35&\h $23.41^{+0.42}_{-0.30}$ \\
SWIFT J1031.9$-$1418&\h &\h 157.9763&\h -14.2809&\h 0.183&\h 0.0851&\h 1&\h 2019-10-30&\h 45.67&\h  8.73&\h H$\alpha$&\h  -1.16&\h 0.93&\h $20.00^{+0.00}_{-0.00}$ \\
SWIFT J1032.7$-$2835&\h ESO 436- G 034&\h 158.1855&\h -28.6101&\h 0.170&\h 0.0121&\h 2&\h 2021-03-12&\h 43.60&\h  8.07&\h $\sigma_*$&\h  -2.57&\h 0.27&\h $22.85^{+0.21}_{-0.14}$ \\
SWIFT J1033.6+7303&\h CGCG 333-038&\h 158.5981&\h  73.0140&\h 0.343&\h 0.0224&\h 1&\h 2021-03-12&\h 44.06&\h  8.01&\h $M_*$&\h  -2.05&\h 0.53&\h $22.62^{+0.17}_{-0.21}$ \\
SWIFT J1038.8$-$4942&\h &\h 159.6883&\h -49.7816&\h 1.360&\h 0.0602&\h 1&\h 2021-03-12&\h 45.15&\h  8.56&\h H$\beta$&\h  -1.51&\h 0.80&\h $22.73^{+0.05}_{-0.05}$ \\
SWIFT J1040.7$-$4619&\h &\h 160.0939&\h -46.4238&\h 0.424&\h 0.0238&\h 2&\h 2019-01-16&\h 44.21&\h  8.52&\h $\sigma_*$&\h  -2.41&\h 0.60&\h $22.60^{+0.03}_{-0.02}$ \\
SWIFT J1042.4+0046&\h &\h 160.5349&\h   0.7017&\h 0.153&\h 0.0952&\h 2&\h 2021-03-12&\h 45.30&\h  7.89&\h $M_*$&\h  -0.69&\h 0.33&\h $22.20^{+0.14}_{-0.12}$ \\
SWIFT J1043.4+1105&\h SDSS J104326.47+110524.2&\h 160.8603&\h  11.0901&\h 0.075&\h 0.0480&\h 1&\h 2021-03-12&\h 44.73&\h  8.29&\h H$\alpha$&\h  -1.66&\h 0.85&\h $20.00^{+0.00}_{-0.00}$ \\
SWIFT J1059.8+6507&\h &\h 164.9315&\h  65.0684&\h 0.070&\h 0.0835&\h 2&\h 2018-07-30&\h 45.16&\h  8.45&\h $\sigma_*$&\h  -1.39&\h 0.79&\h $22.53^{+0.10}_{-0.10}$ \\
SWIFT J1132.9+1019A&\h IC 2921&\h 173.2053&\h  10.2965&\h 0.090&\h 0.0440&\h 1&\h 2019-01-22&\h 44.72&\h  7.84&\h H$\alpha$&\h  -1.22&\h 0.41&\h $21.52^{+0.23}_{-0.22}$ \\
SWIFT J1136.7$-$6007&\h &\h 174.1752&\h -60.0519&\h 2.599&\h 0.0142&\h 2&\h 2019-01-25&\h 43.87&\h  7.92&\h $M_*$&\h  -2.15&\h 0.48&\h $22.43^{+0.15}_{-0.20}$ \\
SWIFT J1139.1+5913&\h SBS 1136+594&\h 174.7870&\h  59.1990&\h 0.041&\h 0.0616&\h 1&\h 2018-03-28&\h 45.14&\h  8.21&\h H$\alpha$&\h  -1.17&\h 0.53&\h $20.00^{+0.00}_{-0.00}$ \\
SWIFT J1143.7+7942&\h UGC 06728&\h 176.3168&\h  79.6815&\h 0.278&\h 0.0063&\h 1&\h 2020-03-02&\h 43.28&\h  6.01&\h H$\alpha$&\h  -0.83&\h 0.78&\h $20.00^{+0.00}_{-0.00}$ \\
SWIFT J1148.3+0901&\h &\h 176.9795&\h   9.0413&\h 0.074&\h 0.0693&\h 1&\h 2019-01-10&\h 45.02&\h  8.63&\h H$\alpha$&\h  -1.71&\h 0.62&\h $20.00^{+0.00}_{-0.00}$ \\
SWIFT J1200.2$-$5350&\h &\h 180.6985&\h -53.8355&\h 0.562&\h 0.0277&\h 2&\h 2020-09-02&\h 44.81&\h  7.98&\h $\sigma_*$&\h  -1.27&\h 0.52&\h $22.46^{+0.02}_{-0.03}$ \\
SWIFT J1211.3$-$3935&\h &\h 182.8095&\h -39.5574&\h 0.233&\h 0.0609&\h 2&\h 2019-01-11&\h 45.16&\h  8.57&\h $\sigma_*$&\h  -1.51&\h 0.49&\h $22.19^{+0.06}_{-0.08}$ \\
SWIFT J1213.1+3239A&\h CGCG 187-022&\h 183.2888&\h  32.5964&\h 0.034&\h 0.0248&\h 2&\h 2018-12-17&\h 44.18&\h  7.75&\h $\sigma_*$&\h  -1.67&\h 0.45&\h $23.74^{+0.29}_{-0.46}$ \\
SWIFT J1217.2$-$2611&\h ESO 505-IG 030&\h 184.2380&\h -26.2093&\h 0.224&\h 0.0393&\h 2&\h 2020-04-24&\h 44.60&\h  7.79&\h $M_*$&\h  -1.29&\h 0.26&\h $23.28^{+0.06}_{-0.20}$ \\
SWIFT J1248.2$-$5828&\h &\h 191.9910&\h -58.5001&\h 1.649&\h 0.0279&\h 2&\h 2020-07-15&\h 44.08&\h  8.03&\h $M_*$&\h  -2.05&\h 0.34&\h $22.46^{+0.23}_{-0.20}$ \\
SWIFT J1306.4$-$4025B&\h &\h 196.8004&\h -40.4076&\h 0.280&\h 0.0159&\h 2&\h 2019-01-13&\h 43.79&\h  7.68&\h $M_*$&\h  -1.99&\h 0.61&\h $20.00^{+0.00}_{-0.00}$ \\
SWIFT J1315.8+4420&\h UGC 08327 NED02&\h 198.8220&\h  44.4071&\h 0.052&\h 0.0355&\h 2&\h 2018-04-23&\h 44.63&\h  8.67&\h $\sigma_*$&\h  -2.14&\h 0.69&\h $22.88^{+0.07}_{-0.05}$ \\
SWIFT J1316.9$-$7155&\h &\h 199.2262&\h -71.9242&\h 0.699&\h 0.0703&\h 1&\h 2018-12-26&\h 45.16&\h  9.03&\h H$\beta$&\h  -1.97&\h 0.86&\h $20.00^{+0.00}_{-0.00}$ \\
SWIFT J1322.2$-$1641&\h MCG -03-34-064&\h 200.6019&\h -16.7286&\h 0.215&\h 0.0168&\h 1&\h 2019-01-18&\h 44.32&\h  7.14&\h H$\alpha$&\h  -0.92&\h 0.69&\h $23.80^{+0.02}_{-0.02}$ \\
SWIFT J1332.0$-$7754&\h &\h 203.1692&\h -77.8446&\h 0.577&\h 0.0098&\h 2&\h 2020-03-30&\h 43.60&\h  8.78&\h $M_*$&\h  -3.28&\h 0.52&\h $23.80^{+0.12}_{-0.08}$ \\
SWIFT J1333.5$-$3401&\h ESO 383-18&\h 203.3587&\h -34.0148&\h 0.163&\h 0.0128&\h 2&\h 2020-04-13&\h 43.85&\h  6.94&\h $M_*$&\h  -1.19&\h 0.41&\h $23.31^{+0.03}_{-0.02}$ \\
SWIFT J1336.0+0304&\h NGC 5231&\h 203.9510&\h   2.9990&\h 0.066&\h 0.0216&\h 2&\h 2019-01-13&\h 44.06&\h  7.95&\h $\sigma_*$&\h  -1.99&\h 0.72&\h $22.34^{+0.04}_{-0.02}$ \\
SWIFT J1338.2+0433&\h NGC 5252&\h 204.5665&\h   4.5426&\h 0.095&\h 0.0229&\h 2&\h 2018-06-18&\h 45.00&\h  8.87&\h $\sigma_*$&\h  -1.97&\h 0.56&\h $22.43^{+0.02}_{-0.02}$ \\
SWIFT J1341.5+6742&\h NGC 5283&\h 205.2740&\h  67.6723&\h 0.052&\h 0.0103&\h 2&\h 2019-10-26&\h 43.23&\h  8.87&\h $\sigma_*$&\h  -3.74&\h 0.85&\h $23.15^{+0.08}_{-0.10}$ \\
SWIFT J1345.5+4139&\h NGC 5290&\h 206.3299&\h  41.7126&\h 0.019&\h 0.0085&\h 2&\h 2018-07-19&\h 43.36&\h  7.76&\h $\sigma_*$&\h  -2.50&\h 0.46&\h $21.96^{+0.08}_{-0.06}$ \\
SWIFT J1349.7+0209&\h UM 614&\h 207.4701&\h   2.0791&\h 0.077&\h 0.0331&\h 1&\h 2019-01-03&\h 44.51&\h  7.50&\h H$\alpha$&\h  -1.09&\h 0.49&\h $21.18^{+0.10}_{-0.07}$ \\
SWIFT J1354.5+1326&\h &\h 208.6211&\h  13.4659&\h 0.077&\h 0.0633&\h 2&\h 2019-01-15&\h 44.89&\h  7.74&\h $M_*$&\h  -0.95&\h 0.38&\h $23.34^{+0.12}_{-0.09}$ \\
SWIFT J1416.9$-$1158&\h &\h 214.2084&\h -11.9829&\h 0.183&\h 0.0992&\h 1&\h 2018-08-15&\h 45.53&\h  9.05&\h H$\beta$&\h  -1.62&\h 0.79&\h $20.00^{+0.00}_{-0.00}$ \\
SWIFT J1421.4+4747&\h SBS 1419+480&\h 215.3742&\h  47.7902&\h 0.048&\h 0.0727&\h 1&\h 2018-08-16&\h 45.27&\h  8.54&\h H$\alpha$&\h  -1.37&\h 0.58&\h $20.00^{+0.00}_{-0.00}$ \\
SWIFT J1424.2+2435&\h NGC 5610&\h 216.0954&\h  24.6144&\h 0.060&\h 0.0169&\h 2&\h 2018-12-18&\h 43.99&\h  7.81&\h $\sigma_*$&\h  -1.92&\h 0.40&\h $22.56^{+0.06}_{-0.10}$ \\
SWIFT J1431.2+2816&\h &\h 217.7700&\h  28.2873&\h 0.057&\h 0.0461&\h 1&\h 2019-01-16&\h 44.52&\h  8.93&\h $M_*$&\h  -2.51&\h 0.73&\h $20.00^{+0.00}_{-0.00}$ \\
SWIFT J1457.8$-$4308&\h IC 4518A&\h 224.4216&\h -43.1321&\h 0.429&\h 0.0158&\h 2&\h 2020-06-14&\h 44.18&\h  6.89&\h $M_*$&\h  -0.81&\h 0.34&\h $23.36^{+0.09}_{-0.08}$ \\
SWIFT J1506.7+0353B&\h &\h 226.6840&\h   3.8620&\h 0.132&\h 0.0373&\h 2&\h 2019-01-11&\h 44.62&\h  7.49&\h $\sigma_*$&\h  -0.97&\h 0.88&\h $22.18^{+0.07}_{-0.08}$ \\
SWIFT J1513.8$-$8125&\h &\h 228.6751&\h -81.3939&\h 0.744&\h 0.0687&\h 1&\h 2020-07-02&\h 45.40&\h  8.71&\h H$\alpha$&\h  -1.41&\h 0.58&\h $20.00^{+0.00}_{-0.00}$ \\
SWIFT J1546.3+6928&\h &\h 236.6014&\h  69.4861&\h 0.114&\h 0.0378&\h 2&\h 2021-03-19&\h 44.62&\h  9.18&\h $\sigma_*$&\h  -2.66&\h 0.27&\h $23.49^{+0.24}_{-0.18}$ \\
SWIFT J1548.5$-$1344&\h NGC 5995&\h 237.1040&\h -13.7578&\h 0.439&\h 0.0244&\h 1&\h 2019-01-18&\h 44.68&\h  7.01&\h H$\alpha$&\h  -0.43&\h 0.82&\h $21.97^{+0.06}_{-0.07}$ \\
SWIFT J1605.9$-$7250&\h &\h 241.3470&\h -72.8990&\h 0.257&\h 0.0900&\h 2&\h 2020-07-08&\h 45.66&\h  9.42&\h $M_*$&\h  -1.86&\h 0.61&\h $23.18^{+0.10}_{-0.10}$ \\
SWIFT J1643.2+7036&\h NGC 6232&\h 250.8343&\h  70.6325&\h 0.121&\h 0.0149&\h 2&\h 2021-05-06&\h 44.15&\h  7.43&\h $\sigma_*$&\h  -1.38&\h 0.85&\h $24.35^{+0.43}_{-0.19}$ \\
SWIFT J1648.0$-$3037&\h &\h 252.0635&\h -30.5845&\h 0.935&\h 0.0310&\h 1&\h 2018-05-27&\h 44.89&\h  7.67&\h $M_*$&\h  -0.88&\h 0.58&\h $21.40^{+0.29}_{-0.50}$ \\
SWIFT J1652.3+5554&\h &\h 253.0780&\h  55.9055&\h 0.053&\h 0.0291&\h 2&\h 2018-08-24&\h 44.37&\h  8.74&\h $M_*$&\h  -2.47&\h 0.36&\h $22.70^{+0.08}_{-0.07}$ \\
SWIFT J1731.3+1442&\h &\h 262.8058&\h  14.7155&\h 0.264&\h 0.0826&\h 1&\h 2021-06-07&\h 45.16&\h  9.00&\h $M_*$&\h  -1.94&\h 0.77&\h $20.00^{+0.00}_{-0.00}$ \\
SWIFT J1737.7$-$5956A&\h &\h 264.4128&\h -59.9407&\h 0.205&\h 0.0171&\h 2&\h 2020-09-02&\h \nd&\h  8.41&\h $M_*$&\h \nd&\h 0.81&\h $-9.00^{+0.00}_{-0.00}$ \\
SWIFT J1741.9$-$1211&\h 2E 1739.1-1210&\h 265.4802&\h -12.1991&\h 1.591&\h 0.0369&\h 1&\h 2020-08-01&\h 45.07&\h  8.04&\h H$\alpha$&\h  -1.07&\h 0.46&\h $21.40^{+0.08}_{-0.08}$ \\
SWIFT J1747.7$-$2253&\h &\h 266.8739&\h -22.8791&\h 2.794&\h 0.0467&\h 1&\h 2018-08-19&\h 44.93&\h  8.99&\h H$\alpha$&\h  -2.16&\h 0.91&\h $22.41^{+0.28}_{-0.37}$ \\
SWIFT J1747.8+6837A&\h Mrk 507&\h 267.1599&\h  68.7044&\h 0.106&\h 0.0551&\h 1&\h 2018-11-14&\h 44.35&\h  7.21&\h H$\alpha$&\h  -0.96&\h 0.64&\h $20.00^{+0.00}_{-0.00}$ \\
SWIFT J1747.8+6837B&\h VII Zw 742&\h 266.7493&\h  68.6102&\h 0.103&\h 0.0630&\h 1&\h 2020-03-15&\h 44.59&\h  7.20&\h H$\alpha$&\h  -0.71&\h 0.86&\h $20.00^{+0.00}_{-0.00}$ \\
SWIFT J1748.8$-$3257&\h &\h 267.2297&\h -32.9145&\h 4.482&\h 0.0200&\h 1&\h 2018-07-06&\h 44.43&\h  7.67&\h H$\alpha$&\h  -1.34&\h 0.61&\h $21.38^{+0.07}_{-0.02}$ \\
SWIFT J1800.3+6637&\h &\h 270.0304&\h  66.6151&\h 0.124&\h 0.0265&\h 2&\h 2018-08-19&\h 44.55&\h  8.88&\h $M_*$&\h  -2.43&\h 0.62&\h $24.02^{+0.23}_{-0.16}$ \\
SWIFT J1807.9+1124&\h &\h 271.9580&\h  11.3470&\h 0.365&\h 0.0787&\h 1&\h 2021-06-04&\h 45.45&\h  8.82&\h H$\alpha$&\h  -1.47&\h 0.73&\h $21.54^{+0.06}_{-0.09}$ \\
SWIFT J1824.2+1845&\h &\h 276.0451&\h  18.7691&\h 0.572&\h 0.0661&\h 1&\h 2021-06-13&\h 45.16&\h  7.88&\h H$\alpha$&\h  -0.82&\h 0.86&\h $22.43^{+0.13}_{-0.12}$ \\
SWIFT J1824.3$-$5624&\h IC 4709&\h 276.0808&\h -56.3692&\h 0.242&\h 0.0167&\h 2&\h 2021-05-28&\h 44.36&\h  8.08&\h $M_*$&\h  -1.82&\h 0.32&\h $23.15^{+0.05}_{-0.07}$ \\
SWIFT J1826.8+3254&\h &\h 276.6350&\h  32.8583&\h 0.294&\h 0.0221&\h 2&\h 2018-08-11&\h 44.22&\h  7.72&\h $\sigma_*$&\h  -1.60&\h 0.31&\h $22.98^{+0.11}_{-0.10}$ \\
SWIFT J1830.8+0928&\h &\h 277.7110&\h   9.4783&\h 0.670&\h 0.0191&\h 2&\h 2020-06-01&\h 43.40&\h  8.39&\h $\sigma_*$&\h  -3.09&\h 0.80&\h $23.18^{+0.22}_{-0.28}$ \\
SWIFT J1844.5$-$6221&\h Fairall 51&\h 281.2249&\h -62.3648&\h 0.297&\h 0.0139&\h 1&\h 2018-07-25&\h 44.15&\h  7.20&\h H$\alpha$&\h  -1.15&\h 0.35&\h $20.00^{+0.00}_{-0.00}$ \\
SWIFT J1845.4+7211&\h &\h 281.3593&\h  72.1838&\h 0.167&\h 0.0463&\h 2&\h 2018-10-02&\h 44.64&\h  7.86&\h $M_*$&\h  -1.32&\h 0.37&\h $22.46^{+0.07}_{-0.10}$ \\
SWIFT J1848.0$-$7832&\h &\h 281.7618&\h -78.5304&\h 0.306&\h 0.0743&\h 1&\h 2019-01-11&\h 45.24&\h  7.73&\h $M_*$&\h  -0.59&\h 0.62&\h $20.00^{+0.00}_{-0.00}$ \\
SWIFT J1856.2$-$7829&\h &\h 284.2823&\h -78.4725&\h 0.427&\h 0.0420&\h 1&\h 2020-07-06&\h 44.88&\h  8.76&\h $M_*$&\h  -1.98&\h 0.55&\h $22.15^{+0.11}_{-0.11}$ \\
SWIFT J1903.9+3349&\h UGC 11397&\h 285.9548&\h  33.8447&\h 0.248&\h 0.0152&\h 2&\h 2019-10-15&\h 43.89&\h  8.24&\h $M_*$&\h  -2.45&\h 0.60&\h $22.87^{+0.11}_{-0.15}$ \\
SWIFT J1905.4+4231&\h &\h 286.3581&\h  42.4610&\h 0.198&\h 0.0279&\h 1&\h 2018-09-24&\h 44.16&\h  8.03&\h $M_*$&\h  -1.97&\h 0.55&\h $20.76^{+0.42}_{-0.30}$ \\
SWIFT J1940.4$-$3015&\h IGR J19405-3016&\h 295.0629&\h -30.2644&\h 0.281&\h 0.0525&\h 1&\h 2020-07-09&\h 45.03&\h  8.37&\h H$\alpha$&\h  -1.44&\h 0.76&\h $20.52^{+0.08}_{-0.11}$ \\
SWIFT J1947.3+4447&\h &\h 296.8307&\h  44.8284&\h 0.535&\h 0.0528&\h 2&\h 2018-12-05&\h 45.12&\h  9.01&\h $\sigma_*$&\h  -1.99&\h 0.75&\h $22.84^{+0.03}_{-0.02}$ \\
SWIFT J1952.4+0237&\h 3C 403&\h 298.0660&\h   2.5070&\h 0.520&\h 0.0584&\h 2&\h 2020-07-19&\h 45.47&\h  9.15&\h $\sigma_*$&\h  -1.78&\h 0.87&\h $23.69^{+0.13}_{-0.04}$ \\
SWIFT J2006.5+5619&\h &\h 301.6389&\h  56.3435&\h 1.013&\h 0.0423&\h 2&\h 2019-01-11&\h 44.66&\h  7.82&\h $M_*$&\h  -1.26&\h 0.22&\h $23.48^{+0.09}_{-0.10}$ \\
SWIFT J2010.7+4801&\h &\h 302.5725&\h  48.0059&\h 1.106&\h 0.0254&\h 2&\h 2018-12-05&\h 44.18&\h  9.48&\h $\sigma_*$&\h  -3.40&\h 0.59&\h $23.00^{+0.17}_{-0.43}$ \\
SWIFT J2018.4$-$5539&\h PKS 2014-55&\h 304.5050&\h -55.6590&\h 0.184&\h 0.0607&\h 2&\h 2020-07-10&\h 45.39&\h  8.50&\h $M_*$&\h  -1.21&\h 0.63&\h $23.45^{+0.10}_{-0.08}$ \\
SWIFT J2018.8+4041&\h &\h 304.6613&\h  40.6834&\h 9.336&\h 0.0144&\h 2&\h 2019-01-17&\h 44.01&\h  7.85&\h $M_*$&\h  -1.94&\h 0.87&\h $22.78^{+0.04}_{-0.06}$ \\
SWIFT J2021.9+4400&\h &\h 305.4544&\h  44.0110&\h 3.620&\h 0.0175&\h 2&\h 2018-12-05&\h 43.85&\h  7.52&\h $M_*$&\h  -1.77&\h 0.52&\h $23.02^{+0.34}_{-0.24}$ \\
SWIFT J2027.1$-$0220&\h &\h 306.7328&\h  -2.2775&\h 0.254&\h 0.0291&\h 2&\h 2018-07-26&\h 44.53&\h  7.83&\h $M_*$&\h  -1.40&\h 0.94&\h $23.82^{+0.15}_{-0.11}$ \\
SWIFT J2035.2+2604&\h &\h 308.7735&\h  26.0583&\h 0.746&\h 0.0478&\h 1&\h 2018-09-30&\h 44.67&\h  7.53&\h H$\alpha$&\h  -0.96&\h 0.26&\h $21.30^{+0.34}_{-0.29}$ \\
SWIFT J2040.2$-$5126&\h ESO 234-IG 063&\h 310.0656&\h -51.4297&\h 0.077&\h 0.0541&\h 2&\h 2021-05-22&\h 44.91&\h  8.11&\h $\sigma_*$&\h  -1.30&\h 0.73&\h $23.41^{+0.20}_{-0.18}$ \\
SWIFT J2044.0+2832&\h RX J2044.0+2833&\h 311.0188&\h  28.5534&\h 0.941&\h 0.0489&\h 1&\h 2018-10-04&\h 44.93&\h  8.23&\h H$\alpha$&\h  -1.40&\h 0.64&\h $21.15^{+0.12}_{-0.09}$ \\
SWIFT J2052.0$-$5704&\h IC 5063&\h 313.0098&\h -57.0688&\h 0.169&\h 0.0115&\h 2&\h 2019-11-25&\h 44.31&\h  8.17&\h $M_*$&\h  -1.96&\h 0.83&\h $23.56^{+0.07}_{-0.01}$ \\
SWIFT J2109.1$-$0942&\h &\h 317.2915&\h  -9.6707&\h 0.584&\h 0.0268&\h 1&\h 2018-09-18&\h 44.40&\h  7.52&\h H$\alpha$&\h  -1.22&\h 0.87&\h $21.20^{+0.32}_{-0.50}$ \\
SWIFT J2114.4+8206&\h &\h 318.5049&\h  82.0801&\h 0.419&\h 0.0833&\h 1&\h 2018-04-02&\h 45.68&\h  9.02&\h H$\alpha$&\h  -1.44&\h 0.82&\h $20.00^{+0.00}_{-0.00}$ \\
SWIFT J2118.9+3336&\h &\h 319.8714&\h  33.5491&\h 0.589&\h 0.0509&\h 1&\h 2019-01-03&\h 44.77&\h  8.11&\h H$\alpha$&\h  -1.44&\h 0.74&\h $21.57^{+0.17}_{-0.17}$ \\
SWIFT J2124.6+5057&\h 4C 50.55&\h 321.1643&\h  50.9735&\h 6.665&\h 0.0151&\h 1&\h 2018-12-13&\h 44.95&\h  6.69&\h H$\alpha$&\h   0.16&\h 0.70&\h $23.02^{+0.08}_{-0.14}$ \\
SWIFT J2127.4+5654&\h &\h 321.9391&\h  56.9430&\h 3.523&\h 0.0147&\h 1&\h 2018-08-19&\h 44.15&\h  6.60&\h $M_*$&\h  -0.55&\h 0.83&\h $20.00^{+0.00}_{-0.00}$ \\
SWIFT J2156.1+4728&\h &\h 323.9750&\h  47.4727&\h 1.741&\h 0.0253&\h 1&\h 2018-12-06&\h 44.39&\h  7.47&\h H$\alpha$&\h  -1.18&\h 0.34&\h $21.28^{+0.24}_{-0.10}$ \\
SWIFT J2201.9$-$3152&\h NGC 7172&\h 330.5080&\h -31.8698&\h 0.073&\h 0.0083&\h 2&\h 2020-09-15&\h 44.22&\h  8.45&\h $\sigma_*$&\h  -2.33&\h 0.70&\h $22.91^{+0.01}_{-0.01}$ \\
SWIFT J2204.7+0337&\h &\h 331.0799&\h   3.5639&\h 0.126&\h 0.0611&\h 2&\h 2020-07-19&\h 45.10&\h  8.25&\h $M_*$&\h  -1.25&\h 0.96&\h $22.83^{+0.15}_{-0.14}$ \\
SWIFT J2214.2$-$2557&\h &\h 333.5382&\h -25.9636&\h 0.065&\h 0.0519&\h 2&\h 2020-09-16&\h 44.89&\h  8.17&\h $\sigma_*$&\h  -1.38&\h 0.83&\h $23.48^{+0.23}_{-0.12}$ \\
SWIFT J2219.7+2614&\h &\h 334.9573&\h  26.2244&\h 0.297&\h 0.0877&\h 1&\h 2019-01-20&\h 45.41&\h  9.01&\h H$\alpha$&\h  -1.70&\h 0.67&\h $21.52^{+0.06}_{-0.04}$ \\
SWIFT J2237.0+2543&\h &\h 339.1370&\h  25.7632&\h 0.143&\h 0.0246&\h 2&\h 2018-04-27&\h 44.07&\h  8.43&\h $M_*$&\h  -2.46&\h 0.91&\h $23.02^{+0.18}_{-0.09}$ \\
SWIFT J2246.0+3941&\h 3C 452&\h 341.4532&\h  39.6877&\h 0.381&\h 0.0811&\h 1&\h 2018-12-08&\h 45.66&\h  6.58&\h H$\alpha$&\h   0.98&\h 0.70&\h $23.76^{+0.07}_{-0.08}$ \\
SWIFT J2320.8+6434&\h &\h 350.1526&\h  64.5125&\h 4.592&\h 0.0717&\h 2&\h 2019-10-23&\h 45.03&\h  7.65&\h $M_*$&\h  -0.72&\h 0.57&\h $22.04^{+0.41}_{-0.19}$ \\
SWIFT J2330.5+7124&\h IGR J23308+7120&\h 352.6571&\h  71.3796&\h 1.791&\h 0.0369&\h 2&\h 2018-12-15&\h 44.40&\h  8.55&\h $M_*$&\h  -2.25&\h 0.95&\h $22.95^{+0.06}_{-0.06}$ \\
SWIFT J2359.3$-$6058&\h &\h 359.7682&\h -60.9165&\h 0.037&\h 0.0963&\h 2&\h 2020-01-14&\h 45.40&\h  8.92&\h $M_*$&\h  -1.62&\h 0.57&\h $23.16^{+0.02}_{-0.02}$ \\
\enddata
\tablecomments{
Col. (1): Object name.
Col. (2): Alternative name.
Col. (3): Right Ascension in J2000 coordinates.
Col. (4): Declination in J2000 coordinates.
Col. (5): Galactic extinction in the {\it V}\ band.
Col. (6): Redshift.
Col. (7): AGN type determined based on the optical spectrum.
Col. (8): Observation date.
Col. (9): Bolometric luminosity inferred from the intrinsic X-ray luminosity,
assuming a bolometric correction of 8 (Ricci et al. 2017).
Col. (10): Black hole mass.
Col. (11): Method to estimate the black hole mass: 
H$\alpha$: virial method with the broad H$\alpha$;
H$\beta$: virial method with the broad H$\beta$;
$\sigma_*$: $M_{\rm BH}-\sigma_*$ relation; 
$M_*$: $M_{\rm BH}-M_*$ relation.
Col. (12): Eddington ratio ($\lambda=L_{\rm bol}/L_{\rm Edd}$).
Col. (13): Axis ratio.
Col. (14): Column density of neutral hydrogen derived from the hard X-ray spectrum (Ricci et al. 2017).
}
\end{deluxetable}
\end{longrotatetable}

\clearpage
\startlongtable
\begin{deluxetable*}{lcccccc}
\tablecolumns{7}
\tablenum{2}
\tabletypesize{\footnotesize}
\tablewidth{0pc}
\tablecaption{Morphology of Host Galaxies}
\tablehead{
\colhead{\h Source Name} &
\colhead{\h Morp.(1)} &
\colhead{\h Morp.(2)} &
\colhead{\h T/S} &
\colhead{\h Major/Minor} &
\colhead{\h Merging Stage} &
\colhead{\h Nucleus} \\
\colhead{\h (1)} &
\colhead{\h (2)} &
\colhead{\h (3)} &
\colhead{\h (4)} &
\colhead{\h (5)} &
\colhead{\h (6)} &
\colhead{\h (7)} 
}
\startdata
SWIFT J0001.0$-$0708&\h S0&\h Sb&\h \nd&\h \nd&\h m1&\h s \\
SWIFT J0001.6$-$7701&\h SB&\h Sc&\h T&\h Minor&\h m2&\h s \\
SWIFT J0003.3+2737&\h SA&\h Sa&\h T&\h Minor&\h m2&\h s \\
SWIFT J0005.0+7021&\h E/S0&\h \nd&\h S&\h \nd&\h \nd&\h s \\
SWIFT J0006.2+2012&\h E&\h \nd&\h \nd&\h Minor&\h m1&\h s \\
SWIFT J0036.3+4540&\h SA&\h Sc&\h \nd&\h \nd&\h m1&\h s \\
SWIFT J0100.9$-$4750&\h SA&\h Sa&\h T&\h Major&\h m3&\h s \\
SWIFT J0123.9$-$5846&\h SB&\h Sb&\h S&\h Minor&\h m2&\h s \\
SWIFT J0128.4+1631&\h SB&\h Sc&\h \nd&\h \nd&\h \nd&\h s \\
SWIFT J0134.1$-$3625&\h E&\h \nd&\h \nd&\h \nd&\h \nd&\h s \\
SWIFT J0157.2+4715&\h SB&\h Sc&\h T&\h Minor&\h m2&\h s \\
SWIFT J0202.4+6824A&\h SB&\h Sb&\h \nd&\h \nd&\h \nd&\h s \\
SWIFT J0202.4+6824B&\h SB&\h Sb&\h \nd&\h \nd&\h \nd&\h s \\
SWIFT J0206.2$-$0019&\h pec.&\h \nd&\h T&\h Major&\h m4&\h s \\
SWIFT J0234.6$-$0848&\h pec.&\h \nd&\h T&\h Major&\h m4&\h d \\
SWIFT J0243.9+5324&\h SB&\h Sb&\h \nd&\h \nd&\h \nd&\h s \\
SWIFT J0333.3+3720&\h S0&\h Sa&\h S&\h Minor&\h m1&\h s \\
SWIFT J0347.0$-$3027&\h S0&\h Sa&\h \nd&\h \nd&\h \nd&\h s \\
SWIFT J0356.9$-$4041&\h E/S0&\h Sa&\h \nd&\h \nd&\h \nd&\h s \\
SWIFT J0405.3$-$3707&\h E/S0&\h Sa&\h \nd&\h \nd&\h \nd&\h s \\
SWIFT J0429.6$-$2114&\h pec.&\h \nd&\h T&\h Major&\h m4&\h s \\
SWIFT J0443.9+2856&\h SB&\h Sb&\h \nd&\h \nd&\h \nd&\h s \\
SWIFT J0446.4+1828&\h SB&\h Sb&\h \nd&\h \nd&\h \nd&\h s \\
SWIFT J0456.3$-$7532&\h SB0&\h Sa&\h \nd&\h \nd&\h \nd&\h s \\
SWIFT J0504.6$-$7345&\h SA&\h Sb&\h \nd&\h \nd&\h \nd&\h s \\
SWIFT J0510.7+1629&\h E/S0&\h \nd&\h \nd&\h \nd&\h \nd&\h s \\
SWIFT J0516.2$-$0009&\h SA&\h Sa&\h \nd&\h \nd&\h \nd&\h s \\
SWIFT J0526.2$-$2118&\h SB0&\h Sa&\h S&\h Minor&\h \nd&\h s \\
SWIFT J0528.1$-$3933&\h SB&\h Sa&\h T&\h Major&\h m2&\h dd \\
SWIFT J0533.9$-$1318&\h S0&\h Sa&\h \nd&\h \nd&\h \nd&\h s \\
SWIFT J0544.4$-$4328&\h E&\h \nd&\h T&\h Minor&\h m2&\h s \\
SWIFT J0552.5+5929&\h pec.&\h \nd&\h T&\h Major&\h m4&\h s \\
SWIFT J0623.9$-$6058&\h SB&\h Sa&\h T&\h Minor&\h m1&\h s \\
SWIFT J0641.3+3257&\h S0&\h Sa&\h T&\h Major&\h m1&\h s \\
SWIFT J0645.9+5303&\h SA&\h Sa&\h \nd&\h \nd&\h \nd&\h s \\
SWIFT J0707.1+6433&\h E/S0&\h Sa&\h S&\h Minor&\h m1&\h s \\
SWIFT J0709.0$-$4642&\h E&\h \nd&\h S&\h Minor&\h m1&\h s \\
SWIFT J0736.9+5846&\h SB&\h Sc&\h \nd&\h \nd&\h \nd&\h s \\
SWIFT J0743.0+6513&\h E/S0&\h Sa&\h \nd&\h \nd&\h \nd&\h s \\
SWIFT J0743.3$-$2546&\h SB&\h Sc&\h \nd&\h \nd&\h \nd&\h s \\
SWIFT J0747.5+6057&\h SB&\h Sb&\h \nd&\h \nd&\h \nd&\h s \\
SWIFT J0747.6$-$7326&\h S0&\h Sa&\h \nd&\h \nd&\h \nd&\h s \\
SWIFT J0753.1+4559&\h E&\h \nd&\h \nd&\h \nd&\h \nd&\h s \\
SWIFT J0756.3$-$4137&\h E&\h \nd&\h \nd&\h \nd&\h \nd&\h s \\
SWIFT J0759.8$-$3844&\h E&\h \nd&\h \nd&\h \nd&\h \nd&\h s \\
SWIFT J0800.1+2638&\h SB&\h Sb&\h \nd&\h \nd&\h m1&\h s \\
SWIFT J0801.9$-$4946&\h SA&\h Sc&\h \nd&\h \nd&\h \nd&\h s \\
SWIFT J0804.6+1045&\h pec.&\h \nd&\h T&\h Major&\h m4&\h s \\
SWIFT J0805.1$-$0110&\h SA&\h Sa&\h \nd&\h \nd&\h \nd&\h s \\
SWIFT J0807.9+3859&\h SB&\h Sa&\h \nd&\h \nd&\h \nd&\h s \\
SWIFT J0819.2$-$2259&\h SB&\h Sa&\h \nd&\h \nd&\h \nd&\h s \\
SWIFT J0823.4$-$0457&\h SA&\h Sa&\h \nd&\h \nd&\h m1&\h s \\
SWIFT J0840.2+2947&\h E/S0&\h Sa&\h S&\h Minor&\h m2&\h s \\
SWIFT J0855.6+6425&\h SA&\h Sa&\h \nd&\h \nd&\h \nd&\h s \\
SWIFT J0902.7$-$4814&\h E&\h \nd&\h \nd&\h \nd&\h \nd&\h s \\
SWIFT J0923.7+2255&\h SB&\h Sb&\h \nd&\h \nd&\h \nd&\h s \\
SWIFT J0924.2$-$3141&\h E&\h \nd&\h \nd&\h \nd&\h \nd&\h s \\
SWIFT J0925.2$-$8423&\h E&\h \nd&\h T&\h Minor&\h m2&\h s \\
SWIFT J0936.2$-$6553&\h SB0&\h Sa&\h \nd&\h \nd&\h \nd&\h s \\
SWIFT J0942.2+2344&\h SB0&\h Sa&\h \nd&\h \nd&\h \nd&\h s \\
SWIFT J0947.6$-$3057&\h SB0&\h Sa&\h \nd&\h \nd&\h \nd&\h s \\
SWIFT J1020.5$-$0237B&\h SB&\h Sb&\h \nd&\h \nd&\h m1&\h s \\
SWIFT J1021.7$-$0327&\h E&\h \nd&\h \nd&\h \nd&\h \nd&\h s \\
SWIFT J1029.8$-$3821&\h SA&\h Sa&\h \nd&\h \nd&\h m1&\h s \\
SWIFT J1031.9$-$1418&\h E&\h \nd&\h \nd&\h \nd&\h m1&\h s \\
SWIFT J1032.7$-$2835&\h S0&\h Sb&\h \nd&\h \nd&\h \nd&\h s \\
SWIFT J1033.6+7303&\h E/S0&\h Sa&\h \nd&\h \nd&\h \nd&\h s \\
SWIFT J1038.8$-$4942&\h E&\h \nd&\h \nd&\h \nd&\h \nd&\h s \\
SWIFT J1040.7$-$4619&\h S0&\h Sa&\h \nd&\h \nd&\h \nd&\h s \\
SWIFT J1042.4+0046&\h SA&\h Sb&\h \nd&\h \nd&\h m1&\h s \\
SWIFT J1043.4+1105&\h E&\h \nd&\h \nd&\h \nd&\h \nd&\h s \\
SWIFT J1059.8+6507&\h pec.&\h \nd&\h T&\h Minor&\h m3&\h s \\
SWIFT J1132.9+1019A&\h SB&\h Sc&\h \nd&\h \nd&\h \nd&\h s \\
SWIFT J1136.7$-$6007&\h S0&\h Sb&\h \nd&\h \nd&\h \nd&\h s \\
SWIFT J1139.1+5913&\h E/S0&\h \nd&\h S&\h Minor&\h m2&\h s \\
SWIFT J1143.7+7942&\h SB0&\h Sa&\h \nd&\h \nd&\h \nd&\h s \\
SWIFT J1148.3+0901&\h SB0&\h Sa&\h \nd&\h \nd&\h m1&\h s \\
SWIFT J1200.2$-$5350&\h pec.&\h \nd&\h S&\h \nd&\h m4&\h s \\
SWIFT J1211.3$-$3935&\h SB&\h Sb&\h \nd&\h \nd&\h \nd&\h s \\
SWIFT J1213.1+3239A&\h S0&\h Sb&\h \nd&\h \nd&\h \nd&\h s \\
SWIFT J1217.2$-$2611&\h SB&\h Sb&\h T&\h Major&\h m2&\h s \\
SWIFT J1248.2$-$5828&\h S0&\h Sb&\h T&\h Major&\h m2&\h s \\
SWIFT J1306.4$-$4025B&\h SB0&\h Sa&\h \nd&\h \nd&\h \nd&\h s \\
SWIFT J1315.8+4420&\h pec.&\h \nd&\h T&\h Major&\h m3&\h s \\
SWIFT J1316.9$-$7155&\h E&\h \nd&\h \nd&\h \nd&\h \nd&\h s \\
SWIFT J1322.2$-$1641&\h SB&\h Sb&\h \nd&\h \nd&\h \nd&\h s \\
SWIFT J1332.0$-$7754&\h SA&\h Sa&\h \nd&\h \nd&\h \nd&\h s \\
SWIFT J1333.5$-$3401&\h SA&\h Sc&\h \nd&\h \nd&\h \nd&\h s \\
SWIFT J1336.0+0304&\h SA&\h Sb&\h \nd&\h \nd&\h \nd&\h s \\
SWIFT J1338.2+0433&\h S0&\h Sa&\h \nd&\h \nd&\h \nd&\h s \\
SWIFT J1341.5+6742&\h SB0&\h Sa&\h \nd&\h \nd&\h m1&\h s \\
SWIFT J1345.5+4139&\h SA&\h Sc&\h \nd&\h \nd&\h \nd&\h s \\
SWIFT J1349.7+0209&\h S0&\h Sa&\h \nd&\h \nd&\h m1&\h s \\
SWIFT J1354.5+1326&\h SA&\h Sb&\h T&\h Major&\h m3&\h s \\
SWIFT J1416.9$-$1158&\h E&\h \nd&\h \nd&\h \nd&\h \nd&\h s \\
SWIFT J1421.4+4747&\h E/S0&\h Sa&\h \nd&\h Minor&\h m1&\h s \\
SWIFT J1424.2+2435&\h SB&\h Sb&\h \nd&\h Minor&\h \nd&\h s \\
SWIFT J1431.2+2816&\h SB&\h Sc&\h \nd&\h \nd&\h m1&\h s \\
SWIFT J1457.8$-$4308&\h pec.&\h \nd&\h T&\h Major&\h m3&\h dd \\
SWIFT J1506.7+0353B&\h SB0&\h Sa&\h \nd&\h \nd&\h \nd&\h s \\
SWIFT J1513.8$-$8125&\h S0&\h Sa&\h T&\h Major&\h m3&\h dd \\
SWIFT J1546.3+6928&\h SB0&\h Sb&\h \nd&\h \nd&\h m1&\h s \\
SWIFT J1548.5$-$1344&\h SB&\h Sc&\h \nd&\h \nd&\h \nd&\h s \\
SWIFT J1605.9$-$7250&\h SA&\h Sb&\h \nd&\h \nd&\h \nd&\h s \\
SWIFT J1643.2+7036&\h SB&\h Sa&\h \nd&\h \nd&\h \nd&\h s \\
SWIFT J1648.0$-$3037&\h E&\h \nd&\h \nd&\h \nd&\h \nd&\h s \\
SWIFT J1652.3+5554&\h SB&\h Sc&\h \nd&\h \nd&\h \nd&\h s \\
SWIFT J1731.3+1442&\h SA&\h Sa&\h T&\h \nd&\h \nd&\h s \\
SWIFT J1737.7$-$5956A&\h SA&\h Sb&\h \nd&\h \nd&\h \nd&\h s \\
SWIFT J1741.9$-$1211&\h SB&\h Sa&\h \nd&\h \nd&\h \nd&\h s \\
SWIFT J1747.7$-$2253&\h E&\h \nd&\h \nd&\h \nd&\h \nd&\h s \\
SWIFT J1747.8+6837A&\h SB&\h Sb&\h \nd&\h \nd&\h m1&\h s \\
SWIFT J1747.8+6837B&\h E&\h \nd&\h T&\h Major&\h m3&\h dd \\
SWIFT J1748.8$-$3257&\h E/S0&\h \nd&\h \nd&\h \nd&\h \nd&\h s \\
SWIFT J1800.3+6637&\h SB&\h Sb&\h \nd&\h \nd&\h \nd&\h s \\
SWIFT J1807.9+1124&\h E&\h \nd&\h \nd&\h \nd&\h \nd&\h s \\
SWIFT J1824.2+1845&\h SA&\h Sa&\h \nd&\h \nd&\h \nd&\h s \\
SWIFT J1824.3$-$5624&\h SA&\h Sa&\h \nd&\h \nd&\h \nd&\h s \\
SWIFT J1826.8+3254&\h S0&\h Sa&\h \nd&\h \nd&\h \nd&\h s \\
SWIFT J1830.8+0928&\h SB0&\h Sa&\h \nd&\h \nd&\h \nd&\h s \\
SWIFT J1844.5$-$6221&\h SB&\h Sb&\h \nd&\h \nd&\h \nd&\h s \\
SWIFT J1845.4+7211&\h pec.&\h \nd&\h T&\h Major&\h m4&\h d \\
SWIFT J1848.0$-$7832&\h pec.&\h \nd&\h T&\h Major&\h m4&\h d \\
SWIFT J1856.2$-$7829&\h S0&\h Sb&\h \nd&\h \nd&\h \nd&\h s \\
SWIFT J1903.9+3349&\h SB&\h Sa&\h \nd&\h \nd&\h \nd&\h s \\
SWIFT J1905.4+4231&\h SB&\h Sa&\h \nd&\h \nd&\h \nd&\h s \\
SWIFT J1940.4$-$3015&\h SB&\h Sc&\h \nd&\h \nd&\h \nd&\h s \\
SWIFT J1947.3+4447&\h SA&\h Sa&\h \nd&\h \nd&\h \nd&\h s \\
SWIFT J1952.4+0237&\h E&\h \nd&\h \nd&\h \nd&\h \nd&\h s \\
SWIFT J2006.5+5619&\h SA&\h Sa&\h \nd&\h \nd&\h \nd&\h s \\
SWIFT J2010.7+4801&\h SB0&\h Sa&\h \nd&\h \nd&\h \nd&\h s \\
SWIFT J2018.4$-$5539&\h E/S0&\h \nd&\h \nd&\h \nd&\h \nd&\h s \\
SWIFT J2018.8+4041&\h SA&\h Sc&\h \nd&\h \nd&\h \nd&\h s \\
SWIFT J2021.9+4400&\h SB0&\h Sa&\h \nd&\h \nd&\h \nd&\h s \\
SWIFT J2027.1$-$0220&\h S0&\h Sa&\h \nd&\h \nd&\h \nd&\h s \\
SWIFT J2035.2+2604&\h SA&\h Sa&\h \nd&\h \nd&\h \nd&\h s \\
SWIFT J2040.2$-$5126&\h pec.&\h \nd&\h T&\h Major&\h m4&\h s \\
SWIFT J2044.0+2832&\h SB&\h Sa&\h \nd&\h \nd&\h \nd&\h s \\
SWIFT J2052.0$-$5704&\h E&\h \nd&\h \nd&\h \nd&\h \nd&\h s \\
SWIFT J2109.1$-$0942&\h SB&\h Sa&\h \nd&\h \nd&\h \nd&\h s \\
SWIFT J2114.4+8206&\h E&\h \nd&\h \nd&\h \nd&\h \nd&\h s \\
SWIFT J2118.9+3336&\h E&\h \nd&\h S&\h Minor&\h m1&\h s \\
SWIFT J2124.6+5057&\h E&\h \nd&\h \nd&\h \nd&\h \nd&\h s \\
SWIFT J2127.4+5654&\h E&\h \nd&\h \nd&\h \nd&\h \nd&\h s \\
SWIFT J2156.1+4728&\h SA&\h Sc&\h \nd&\h \nd&\h \nd&\h s \\
SWIFT J2201.9$-$3152&\h E/S0&\h \nd&\h \nd&\h \nd&\h \nd&\h s \\
SWIFT J2204.7+0337&\h pec.&\h \nd&\h T&\h Minor&\h m4&\h s \\
SWIFT J2214.2$-$2557&\h SB&\h Sb&\h \nd&\h \nd&\h \nd&\h s \\
SWIFT J2219.7+2614&\h E&\h \nd&\h S&\h Minor&\h m1&\h s \\
SWIFT J2237.0+2543&\h SB&\h Sb&\h \nd&\h \nd&\h \nd&\h s \\
SWIFT J2246.0+3941&\h E&\h \nd&\h \nd&\h \nd&\h \nd&\h s \\
SWIFT J2320.8+6434&\h E&\h \nd&\h \nd&\h \nd&\h \nd&\h s \\
SWIFT J2330.5+7124&\h SB&\h Sa&\h \nd&\h \nd&\h \nd&\h s \\
SWIFT J2359.3$-$6058&\h E&\h \nd&\h \nd&\h \nd&\h \nd&\h s \\
\enddata
\tablecomments{
Col. (1): Object name.
Col. (2): Morphological classification: E;E/S0;S0;SB0;SA;SB;peculiar.
Col. (3): Morphological classification based on the bulge dominance: Sa/Sb/Sc.
Col. (4): Presence of perturbed structure: ``T" = tidal tail; ``S" = shell structure.
Col. (5): Merging feature: ``Major" = major merger; ``Minor" = minor merger.
Col. (6): Merging stage: 
``m1" = a companion galaxy with very faint (or no) sign of interaction; 
``m2" = a sign of interaction (tail and shell) but no major disturbance in the 
host galaxy;
``m3" = major disturbance in the host galaxy but the companion galaxy is not 
yet merged.
``m4" = two galaxies share the common envelope (latest stage of the merger).
Col. (7): Number of nuclei: ``s" = single nucleus; ``d" = double nuclei in a single host;
``dd" = double nuclei in separate hosts.
}
\end{deluxetable*}

\startlongtable
\begin{longrotatetable}
\begin{deluxetable}{crccccccccccccccc}
\tablecolumns{17}
\tablenum{3}
\tabletypesize{\scriptsize}
\tablewidth{0pc}
\tablecaption{Merging Fraction}
\tablehead{
\colhead{\h} &
\colhead{\h} &
\multicolumn{3}{c}{Tidal/shell} &
\colhead{\h} &
\multicolumn{3}{c}{Major/minor} &
\colhead{\h} &
\multicolumn{3}{c}{m2+m3+m4} &
\colhead{\h} &
\multicolumn{3}{c}{Double nuclei} \\
\cline{3-5} 
\cline{7-9} 
\cline{11-13} 
\cline{15-17} 
\colhead{\h Parameter} &
\colhead{\h Bin} &
\colhead{\h All} &
\colhead{\h Type 1} &
\colhead{\h Type 2} &
\colhead{\h} &
\colhead{\h All} &
\colhead{\h Type 1} &
\colhead{\h Type 2} &
\colhead{\h} &
\colhead{\h All} &
\colhead{\h Type 1} &
\colhead{\h Type 2} &
\colhead{\h} &
\colhead{\h All} &
\colhead{\h Type 1} &
\colhead{\h Type 2} \\
\colhead{\h (1)} &
\colhead{\h (2)} &
\colhead{\h (3)} &
\colhead{\h (4)} &
\colhead{\h (5)} &
\colhead{\h} &
\colhead{\h (6)} &
\colhead{\h (7)} &
\colhead{\h (8)} &
\colhead{\h} &
\colhead{\h (9)} &
\colhead{\h (10)} &
\colhead{\h (11)} &
\colhead{\h} &
\colhead{\h (12)} &
\colhead{\h (13)} &
\colhead{\h (14)} 
}
\startdata
log $L_{\rm bol}$&\h 43.25&\h $0.00^{+0.17}_{-0.00}$&\h $0.00^{+0.54}_{-0.00}$&\h $0.00^{+0.21}_{-0.00}$&\h &\h $0.00^{+0.17}_{-0.00}$&\h $0.00^{+0.54}_{-0.00}$&\h $0.00^{+0.21}_{-0.00}$&\h &\h $0.00^{+0.17}_{-0.00}$&\h $0.00^{+0.54}_{-0.00}$&\h $0.00^{+0.21}_{-0.00}$&\h &\h $0.00^{+0.17}_{-0.00}$&\h $0.00^{+0.54}_{-0.00}$&\h $0.00^{+0.21}_{-0.00}$ \\
&\h 43.75&\h $0.00^{+0.07}_{-0.00}$&\h $0.00^{+0.35}_{-0.00}$&\h $0.00^{+0.08}_{-0.00}$&\h &\h $0.08^{+0.11}_{-0.04}$&\h $0.00^{+0.35}_{-0.00}$&\h $0.09^{+0.12}_{-0.05}$&\h &\h $0.00^{+0.07}_{-0.00}$&\h $0.00^{+0.35}_{-0.00}$&\h $0.00^{+0.08}_{-0.00}$&\h &\h $0.00^{+0.07}_{-0.00}$&\h $0.00^{+0.35}_{-0.00}$&\h $0.00^{+0.08}_{-0.00}$ \\
&\h 44.25&\h $0.13^{+0.06}_{-0.04}$&\h $0.00^{+0.06}_{-0.00}$&\h $0.22^{+0.10}_{-0.07}$&\h &\h $0.15^{+0.07}_{-0.05}$&\h $0.06^{+0.09}_{-0.04}$&\h $0.22^{+0.10}_{-0.07}$&\h &\h $0.10^{+0.06}_{-0.04}$&\h $0.00^{+0.06}_{-0.00}$&\h $0.17^{+0.09}_{-0.06}$&\h &\h $0.05^{+0.05}_{-0.02}$&\h $0.00^{+0.06}_{-0.00}$&\h $0.09^{+0.08}_{-0.04}$ \\
&\h 44.75&\h $0.33^{+0.07}_{-0.06}$&\h $0.21^{+0.08}_{-0.06}$&\h $0.48^{+0.10}_{-0.10}$&\h &\h $0.31^{+0.07}_{-0.06}$&\h $0.21^{+0.08}_{-0.06}$&\h $0.43^{+0.10}_{-0.10}$&\h &\h $0.27^{+0.07}_{-0.06}$&\h $0.17^{+0.08}_{-0.06}$&\h $0.39^{+0.10}_{-0.09}$&\h &\h $0.04^{+0.04}_{-0.02}$&\h $0.03^{+0.05}_{-0.02}$&\h $0.04^{+0.06}_{-0.03}$ \\
&\h 45.25&\h $0.41^{+0.08}_{-0.08}$&\h $0.52^{+0.10}_{-0.10}$&\h $0.25^{+0.12}_{-0.09}$&\h &\h $0.38^{+0.08}_{-0.07}$&\h $0.48^{+0.10}_{-0.10}$&\h $0.25^{+0.12}_{-0.09}$&\h &\h $0.26^{+0.07}_{-0.06}$&\h $0.30^{+0.10}_{-0.09}$&\h $0.19^{+0.11}_{-0.08}$&\h &\h $0.08^{+0.05}_{-0.03}$&\h $0.13^{+0.09}_{-0.05}$&\h $0.00^{+0.06}_{-0.00}$ \\
&\h 45.75&\h $0.00^{+0.17}_{-0.00}$&\h $0.00^{+0.21}_{-0.00}$&\h $0.00^{+0.54}_{-0.00}$&\h &\h $0.00^{+0.17}_{-0.00}$&\h $0.00^{+0.21}_{-0.00}$&\h $0.00^{+0.54}_{-0.00}$&\h &\h $0.00^{+0.17}_{-0.00}$&\h $0.00^{+0.21}_{-0.00}$&\h $0.00^{+0.54}_{-0.00}$&\h &\h $0.00^{+0.17}_{-0.00}$&\h $0.00^{+0.21}_{-0.00}$&\h $0.00^{+0.54}_{-0.00}$ \\
\hline
log $\lambda$&\h -3.50&\h $0.00^{+0.15}_{-0.00}$&\h $0.00^{+0.00}_{-0.00}$&\h $0.00^{+0.15}_{-0.00}$&\h &\h $0.00^{+0.15}_{-0.00}$&\h $0.00^{+0.00}_{-0.00}$&\h $0.00^{+0.15}_{-0.00}$&\h &\h $0.00^{+0.15}_{-0.00}$&\h $0.00^{+0.00}_{-0.00}$&\h $0.00^{+0.15}_{-0.00}$&\h &\h $0.00^{+0.15}_{-0.00}$&\h $0.00^{+0.00}_{-0.00}$&\h $0.00^{+0.15}_{-0.00}$ \\
&\h -2.50&\h $0.16^{+0.09}_{-0.06}$&\h $0.14^{+0.18}_{-0.08}$&\h $0.17^{+0.10}_{-0.07}$&\h &\h $0.16^{+0.09}_{-0.06}$&\h $0.14^{+0.18}_{-0.08}$&\h $0.17^{+0.10}_{-0.07}$&\h &\h $0.12^{+0.08}_{-0.05}$&\h $0.14^{+0.18}_{-0.08}$&\h $0.11^{+0.09}_{-0.05}$&\h &\h $0.00^{+0.04}_{-0.00}$&\h $0.00^{+0.13}_{-0.00}$&\h $0.00^{+0.05}_{-0.00}$ \\
&\h -1.50&\h $0.31^{+0.69}_{-****}$&\h $0.30^{+0.07}_{-0.06}$&\h $0.31^{+0.07}_{-0.06}$&\h &\h $0.32^{+0.68}_{-****}$&\h $0.32^{+0.07}_{-0.06}$&\h $0.31^{+0.07}_{-0.06}$&\h &\h $0.22^{+0.78}_{-****}$&\h $0.19^{+0.06}_{-0.05}$&\h $0.25^{+0.07}_{-0.06}$&\h &\h $0.04^{+0.96}_{-****}$&\h $0.04^{+0.04}_{-0.02}$&\h $0.04^{+0.04}_{-0.02}$ \\
&\h -0.50&\h $0.21^{+0.09}_{-0.07}$&\h $0.17^{+0.10}_{-0.07}$&\h $0.33^{+0.20}_{-0.15}$&\h &\h $0.17^{+0.09}_{-0.06}$&\h $0.11^{+0.09}_{-0.05}$&\h $0.33^{+0.20}_{-0.15}$&\h &\h $0.17^{+0.09}_{-0.06}$&\h $0.11^{+0.09}_{-0.05}$&\h $0.33^{+0.20}_{-0.15}$&\h &\h $0.12^{+0.08}_{-0.05}$&\h $0.11^{+0.09}_{-0.05}$&\h $0.17^{+0.20}_{-0.10}$ \\
&\h  0.50&\h $0.00^{+0.35}_{-0.00}$&\h $0.00^{+0.35}_{-0.00}$&\h $0.00^{+0.00}_{-0.00}$&\h &\h $0.00^{+0.35}_{-0.00}$&\h $0.00^{+0.35}_{-0.00}$&\h $0.00^{+0.00}_{-0.00}$&\h &\h $0.00^{+0.35}_{-0.00}$&\h $0.00^{+0.35}_{-0.00}$&\h $0.00^{+0.00}_{-0.00}$&\h &\h $0.00^{+0.35}_{-0.00}$&\h $0.00^{+0.35}_{-0.00}$&\h $0.00^{+0.00}_{-0.00}$ \\
&\h  1.50&\h $0.00^{+0.54}_{-0.00}$&\h $0.00^{+0.54}_{-0.00}$&\h $0.00^{+0.00}_{-0.00}$&\h &\h $0.00^{+0.54}_{-0.00}$&\h $0.00^{+0.54}_{-0.00}$&\h $0.00^{+0.00}_{-0.00}$&\h &\h $0.00^{+0.54}_{-0.00}$&\h $0.00^{+0.54}_{-0.00}$&\h $0.00^{+0.00}_{-0.00}$&\h &\h $0.00^{+0.54}_{-0.00}$&\h $0.00^{+0.54}_{-0.00}$&\h $0.00^{+0.00}_{-0.00}$ \\
\hline
log $N_{\rm H}$&\h 20.00&\h $0.32^{+0.08}_{-0.07}$&\h $0.33^{+0.08}_{-0.07}$&\h $0.00^{+0.54}_{-0.00}$&\h &\h $0.35^{+0.08}_{-0.07}$&\h $0.36^{+0.08}_{-0.07}$&\h $0.00^{+0.54}_{-0.00}$&\h &\h $0.27^{+0.08}_{-0.07}$&\h $0.28^{+0.08}_{-0.07}$&\h $0.00^{+0.54}_{-0.00}$&\h &\h $0.08^{+0.06}_{-0.03}$&\h $0.08^{+0.06}_{-0.03}$&\h $0.00^{+0.54}_{-0.00}$ \\
&\h 20.50&\h $0.30^{+0.07}_{-0.06}$&\h $0.31^{+0.07}_{-0.07}$&\h $0.00^{+0.54}_{-0.00}$&\h &\h $0.33^{+0.07}_{-0.07}$&\h $0.33^{+0.08}_{-0.07}$&\h $0.00^{+0.54}_{-0.00}$&\h &\h $0.26^{+0.07}_{-0.06}$&\h $0.26^{+0.07}_{-0.06}$&\h $0.00^{+0.54}_{-0.00}$&\h &\h $0.09^{+0.05}_{-0.03}$&\h $0.10^{+0.05}_{-0.04}$&\h $0.00^{+0.54}_{-0.00}$ \\
&\h 21.00&\h $0.12^{+0.10}_{-0.06}$&\h $0.12^{+0.10}_{-0.06}$&\h $0.00^{+0.00}_{-0.00}$&\h &\h $0.12^{+0.10}_{-0.06}$&\h $0.12^{+0.10}_{-0.06}$&\h $0.00^{+0.00}_{-0.00}$&\h &\h $0.06^{+0.08}_{-0.03}$&\h $0.06^{+0.08}_{-0.03}$&\h $0.00^{+0.00}_{-0.00}$&\h &\h $0.06^{+0.08}_{-0.03}$&\h $0.06^{+0.08}_{-0.03}$&\h $0.00^{+0.00}_{-0.00}$ \\
&\h 21.50&\h $0.20^{+0.10}_{-0.07}$&\h $0.19^{+0.11}_{-0.08}$&\h $0.25^{+0.25}_{-0.14}$&\h &\h $0.20^{+0.10}_{-0.07}$&\h $0.19^{+0.11}_{-0.08}$&\h $0.25^{+0.25}_{-0.14}$&\h &\h $0.05^{+0.07}_{-0.03}$&\h $0.00^{+0.06}_{-0.00}$&\h $0.25^{+0.25}_{-0.14}$&\h &\h $0.05^{+0.07}_{-0.03}$&\h $0.00^{+0.06}_{-0.00}$&\h $0.25^{+0.25}_{-0.14}$ \\
&\h 22.00&\h $0.25^{+0.08}_{-0.07}$&\h $0.21^{+0.13}_{-0.09}$&\h $0.28^{+0.11}_{-0.09}$&\h &\h $0.22^{+0.08}_{-0.06}$&\h $0.21^{+0.13}_{-0.09}$&\h $0.22^{+0.11}_{-0.08}$&\h &\h $0.16^{+0.07}_{-0.05}$&\h $0.07^{+0.10}_{-0.04}$&\h $0.22^{+0.11}_{-0.08}$&\h &\h $0.06^{+0.06}_{-0.03}$&\h $0.00^{+0.07}_{-0.00}$&\h $0.11^{+0.09}_{-0.05}$ \\
&\h 22.50&\h $0.26^{+0.07}_{-0.06}$&\h $0.17^{+0.13}_{-0.08}$&\h $0.29^{+0.08}_{-0.07}$&\h &\h $0.24^{+0.07}_{-0.06}$&\h $0.08^{+0.11}_{-0.05}$&\h $0.29^{+0.08}_{-0.07}$&\h &\h $0.22^{+0.07}_{-0.05}$&\h $0.08^{+0.11}_{-0.05}$&\h $0.26^{+0.08}_{-0.07}$&\h &\h $0.02^{+0.03}_{-0.01}$&\h $0.00^{+0.08}_{-0.00}$&\h $0.03^{+0.04}_{-0.02}$ \\
&\h 23.00&\h $0.31^{+0.07}_{-0.06}$&\h $0.20^{+0.22}_{-0.12}$&\h $0.32^{+0.07}_{-0.07}$&\h &\h $0.31^{+0.07}_{-0.06}$&\h $0.00^{+0.17}_{-0.00}$&\h $0.34^{+0.07}_{-0.07}$&\h &\h $0.22^{+0.06}_{-0.05}$&\h $0.00^{+0.17}_{-0.00}$&\h $0.25^{+0.07}_{-0.06}$&\h &\h $0.02^{+0.03}_{-0.01}$&\h $0.00^{+0.17}_{-0.00}$&\h $0.02^{+0.03}_{-0.01}$ \\
&\h 23.50&\h $0.23^{+0.07}_{-0.06}$&\h $0.00^{+0.17}_{-0.00}$&\h $0.26^{+0.08}_{-0.07}$&\h &\h $0.23^{+0.07}_{-0.06}$&\h $0.00^{+0.17}_{-0.00}$&\h $0.26^{+0.08}_{-0.07}$&\h &\h $0.15^{+0.07}_{-0.05}$&\h $0.00^{+0.17}_{-0.00}$&\h $0.18^{+0.07}_{-0.06}$&\h &\h $0.03^{+0.04}_{-0.01}$&\h $0.00^{+0.17}_{-0.00}$&\h $0.03^{+0.04}_{-0.02}$ \\
&\h 24.00&\h $0.06^{+0.08}_{-0.03}$&\h $0.00^{+0.26}_{-0.00}$&\h $0.07^{+0.09}_{-0.04}$&\h &\h $0.06^{+0.08}_{-0.03}$&\h $0.00^{+0.26}_{-0.00}$&\h $0.07^{+0.09}_{-0.04}$&\h &\h $0.06^{+0.08}_{-0.03}$&\h $0.00^{+0.26}_{-0.00}$&\h $0.07^{+0.09}_{-0.04}$&\h &\h $0.00^{+0.05}_{-0.00}$&\h $0.00^{+0.26}_{-0.00}$&\h $0.00^{+0.06}_{-0.00}$ \\
&\h 24.50&\h $0.00^{+0.17}_{-0.00}$&\h $0.00^{+0.00}_{-0.00}$&\h $0.00^{+0.17}_{-0.00}$&\h &\h $0.00^{+0.17}_{-0.00}$&\h $0.00^{+0.00}_{-0.00}$&\h $0.00^{+0.17}_{-0.00}$&\h &\h $0.00^{+0.17}_{-0.00}$&\h $0.00^{+0.00}_{-0.00}$&\h $0.00^{+0.17}_{-0.00}$&\h &\h $0.00^{+0.17}_{-0.00}$&\h $0.00^{+0.00}_{-0.00}$&\h $0.00^{+0.17}_{-0.00}$ \\
\enddata
\tablecomments{
Col. (1): Parameter.
Col. (2): Bin. The units for $L_{\rm bol}$ and $N_{\rm H}$ are
erg s$^{-1}$ and cm$^{-2}$, respectively. 
Col. (3): Merging fraction of the entire sample, based on the presence of tidal tail or shell.
Col. (4): Merging fraction of type 1 AGN, based on the presence of tidal tail or shell.
Col. (5): Merging fraction of type 2 AGN, based on the presence of tidal tail or shell.
Col. (6): Merging fraction of the entire sample, based on the presence of major/minor merging features.
Col. (7): Merging fraction of type 1 AGN, based on the presence of major/minor merging features.
Col. (8): Merging fraction of type 2 AGN, based on the presence of major/minor merging features.
Col. (9): Merging fraction of the entire sample, based on the merging stage (m2+m3+m4).
Col. (10): Merging fraction of type 1 AGN, based on the merging stage (m2+m3+m4).
Col. (11): Merging fraction of type 2 AGN, based on the merging stage (m2+m3+m4).
Col. (12): Fraction of the entire sample, based on the presence of double nuclei.
Col. (13): Fraction of type 1 AGN, based on the presence of double nuclei.
Col. (14): Fraction of type 2 AGN, based on the presence of double nuclei.
}
\end{deluxetable}
\end{longrotatetable}

\end{document}